\documentclass[useAMS,usenatbib]{mn2e}

\usepackage{graphicx,subfigure,color}
\usepackage[total={17.8cm,24.0cm},centering]{geometry}
\usepackage{times}

\def\hi{H{\small I}}

\def\atlas{{{ATLAS}}$^{\rm 3D}$}
\def\kms{km~s$^{-1}$}
\def\co{CO}

\def\lcdm{$\Lambda$CDM}

\def\arcsec{$^{\prime \prime}$}
\definecolor{Mygrey}{gray}{0.75}

\newcommand{\ltsimeq}{\raisebox{-0.6ex}{$\,\stackrel{\raisebox{-.2ex}{$\textstyle <$}}{\sim}\,$}}
\newcommand{\gtsimeq}{\raisebox{-0.6ex}{$\,\stackrel{\raisebox{-.2ex}{$\textstyle >$}}{\sim}\,$}}

\usepackage[compact]{titlesec}
\titlespacing{\section}{0pt}{*2}{*1}

\title[On the origin of the gas in early-type galaxies]{The \atlas\ project -- X. On the origin of the molecular and ionised gas in early-type galaxies}
\author[Timothy A. Davis et al.]{Timothy A. Davis$^1$\thanks{Email: timothy.davis@astro.ox.ac.uk},
Katherine Alatalo$^2$,
Marc Sarzi$^{3}$, 
Martin Bureau$^1$,
Lisa M. Young$^4$\thanks{Adjunct Astronomer with NRAO},
\newauthor 
Leo Blitz$^2$,
Paolo Serra$^{5}$,
Alison F. Crocker$^{6}$,
Davor Krajnovi\'c$^7$,  
 Richard M.\,McDermid$^{8}$, 
 \newauthor 
  Maxime Bois$^{9}$, 
  Fr\'ed\'eric Bournaud$^{10}$, 
Michele Cappellari$^1$, 
 Roger L. Davies$^1$, 
 \newauthor 
  Pierre-Alain Duc$^{10}$
    P. Tim de Zeeuw$^{7,11}$, 
  Eric Emsellem$^{7,12}$, 
   Sadegh Khochfar$^{13}$, 
   \newauthor 
    Harald Kuntschner$^{14}$, 
      Pierre-Yves Lablanche$^{12}$, 
    Raffaella Morganti$^{5,15}$, 
  Thorsten Naab$^{16}$, 
  \newauthor 
      Tom Oosterloo$^{5,15}$, 
   Nicholas Scott$^1$,
  and Anne-Marie Weijmans$^{17}$\thanks{Dunlap Fellow}\\
$^1$Sub-department of Astrophysics, Department of Physics, University of Oxford, Denys Wilkinson Building, Keble Road, Oxford, OX1 3RH, UK\\
$^2$Department of Astronomy, Campbell Hall, University of California, Berkeley, CA 94720, USA\\
$^{3}$Centre for Astrophysics Research, University of Hertfordshire, Hatfield, Herts AL1 9AB, UK\\
$^4$Physics Department, New Mexico Institute of Mining and Technology, Socorro, NM 87801, USA\\
$^{5}$Netherlands Institute for Radio Astronomy (ASTRON), Postbus 2, 7990 AA Dwingeloo, The Netherlands\\
$^{6}$Department of Astronomy, University of Massachusetts, Amherst, MA 01003, USA\\
$^7$European Southern Observatory, Karl-Schwarzschild-Str. 2, 85748 Garching, Germany\\
$^{8}$Gemini Observatory, Northern Operations Centre, 670 N. A`ohoku Place, Hilo, HI 96720, USA\\
$^{9}$Observatoire de Paris, LERMA, 61 Av. de l`Observatoire, 75014, Paris, France\\
$^{10}$Laboratoire AIM Paris-Saclay, CEA/IRFU/SAp Ð CNRS Ð Universit\'e Paris Diderot, 91191 Gif-sur-Yvette Cedex, France\\
$^{11}$Sterrewacht Leiden, Leiden University, Postbus 9513, 2300 RA Leiden, the Netherlands\\
$^{12}$Universit\'e Lyon 1, Observatoire de Lyon, Centre de Recherche Astrophysique de Lyon and Ecole Nationale Sup\'erieure de Lyon,\\\,\, \,9 avenue Charles Andr\'e, F-69230 Saint-Genis Laval, France\\
$^{13}$Max-Planck Institut f\"ur extraterrestrische Physik, PO Box 1312, D-85478 Garching, Germany\\
$^{14}$Space Telescope European Coordinating Facility, European Southern Observatory, Karl-Schwarzschild-Str. 2, 85748 Garching, Germany\\
$^{15}$Kapteyn Astronomical Institute, University of Groningen, Postbus 800, 9700 AV Groningen, The Netherlands\\
$^{16}$Max-Planck-Institut f\"ur Astrophysik, Karl-Schwarzschild-Str. 1, 85741 Garching, Germany\\ 
$^{17}$Dunlap Institute for Astronomy \& Astrophysics, University of Toronto, 50 St. George Street, Toronto, ON M5S 3H4, Canada 
}
\begin{document}
\date{Accepted 2011 June 29. Received 2011 June 28; in original form 2011 March 28}
\pagerange{\pageref{firstpage}--\pageref{lastpage}} \pubyear{2009}
\maketitle
\label{firstpage}
\begin{abstract}

We make use of interferometric CO and \hi\ observations, and optical integral-field spectroscopy from the \atlas\ survey to probe the origin of the molecular and ionised interstellar medium (ISM) in local early-type galaxies. We find that 36$\pm$5\% of our sample of fast rotating early-type galaxies have their ionised gas kinematically misaligned with respect to the stars, setting a strong lower limit on the importance of externally acquired gas (e.g. from mergers and cold accretion). Slow rotators have a flat distribution of misalignments, indicating that the dominant source of gas is external. The molecular, ionised and atomic gas in all the detected galaxies are always kinematically aligned, even when they are misaligned from the stars, suggesting that all these three phases of the interstellar medium share a common origin. 
In addition, we find that the origin of the cold and warm gas in fast-rotating early-type galaxies is strongly affected by environment, despite the molecular gas detection rate and mass fractions being fairly independent of group/cluster membership. Galaxies in dense groups and the Virgo cluster  nearly always have their molecular gas kinematically aligned with the stellar kinematics, consistent with a purely internal origin (presumably stellar mass loss). In the field, however, kinematic misalignments between the stellar and gaseous components indicate that at least 42$\pm$5\% of local fast-rotating early-type galaxies have their gas supplied from external sources. When one also considers evidence of accretion present in the galaxies' atomic gas distributions, $\gtsimeq$46\% of fast-rotating field ETGs are likely to have acquired a detectable amount of ISM from accretion and mergers. We discuss several scenarios which could explain the environmental dichotomy, including preprocessing in galaxy groups/cluster outskirts and the morphological transformation of spiral galaxies, but we find it difficult to simultaneously explain the kinematic misalignment difference and the constant detection rate.
Furthermore, our results suggest that galaxy mass may be an important independent factor associated with the origin of the gas, with the most massive fast-rotating galaxies in our sample (M$_K\ltsimeq-24$ mag; stellar mass of $\approx$8$\times$10$^{10}$ M$_{\odot}$) always having kinematically aligned gas. This mass dependence appears to be independent of environment, suggesting it is caused by a separate physical mechanism.

\end{abstract}


\section{Introduction}
Lenticular and elliptical galaxies, collectively known as early-type galaxies (ETGs), have long been thought to be the endpoint of galaxy evolution. These systems have surprisingly uniform red optical colours and are located in a tight red sequence in an optical colour-magnitude diagram \citep[e.g.][]{Baldry:2004p3398}.  
The origin of the colour bimodality that separates the ETGs in the red sequence from star-forming galaxies, located in a `blue cloud', has become a central question in extragalactic astrophysics.  In order for galaxies to fade and join this tight red sequence rapidly enough, it is thought that the fuel for star formation must be consumed, destroyed or removed on a reasonably short timescale \citep[e.g.][]{Faber:2007p3453}.

 In the paradigm of cold dark matter with a non-zero cosmological constant (\lcdm), it is thought that star-formation is shut-off in these galaxies by major mergers that can funnel the cold gas into the centre of the remnant \citep{Barnes:2002p2041}, where it may be destroyed by an active galactic nucleus (AGN) or be used up quickly in a starburst.
In massive galaxies, virial shocks \citep[such as those discussed by][]{Birnboim:2003p3427} can also create a halo mass threshold, above which incoming gas will always shock to the virial temperature, stopping the accretion of further cold gas. Energy input from an AGN or starburst may also remove cold gas without a major merger, and keep halo gas from cooling. Galaxies in clusters and groups may also have their interstellar medium (ISM) removed by harassment, or stripped and ablated away by a hot intra-cluster medium (ICM). 

Evolutionary processes like those listed above leave galaxies on the red sequence with little or no cold ISM, and thus no star formation. Evidence is mounting, however, that a reasonable fraction of ETGs do have cold gas reservoirs and residual star-formation. For example 
the surveys of \cite{Grossi:2009p3448}, \cite{Morganti:2006p1934} and \cite{Oosterloo:2010p3376} find between 44 and 66\% of field ETGs have detectable neutral hydrogen reservoirs, while
\cite{Colbert:2001p1791} have shown that over 75\% of early-types contain dust features in optical images. \cite{Yi:2005p3450} and \cite{Kaviraj:2007p1804} have also shown that at least 30\% of ETGs have an excess of UV emission, attributable to ongoing residual star formation (over and above that predicted for an old stellar population from the UV-upturn phenomenon; see \citealt{Yi:2008p3435} for a review). \cite{Sarzi:2006p1474} have shown that at least 10\% of local ETGs have strong Balmer line emission and line ratios indicative of ongoing star-formation. These studies all imply some ETGs must have a cold molecular gas reservoir to fuel this ongoing star-formation. 

The molecular gas content of early-types had until recently been investigated mainly using infrared-selected surveys. For example \cite{Knapp:1996p1859} report a molecular gas detection rate of 80\% for ETGs brighter than 1 Jy at 100 $\mu$m. \cite{Combes:2007p231} performed one of the first studies of ETGs which was not infrared-selected. They observed the SAURON sample of galaxies {(a representative sample of 48 ETGs which had been observed with the SAURON integral field spectrograph, see \citealt{deZeeuw:2002p1496} for full details)} and reported a molecular gas detection rate of 28\% (with a CO(1-0) sensitivity of 3mK T$_a^{*}$ in a 30 \kms\ channel). {The survey of \cite{Welch:2010p3290} \citep[which combined and added to the results of][]{Welch:2003p2521,Sage:2006p3466,Sage:2007p3467}} was the first published volume-limited sample, which found a detection rate of 26\% for ETGs within 20 Mpc (with a similar sensitivity).

The \atlas\ project is a complete, volume-limited survey of the properties of 260 morphologically-selected ETGs within 40 Mpc \citep[for full details see][hereafter Paper I]{Cap2010}. This survey combines integral-field spectroscopy (IFS), optical imaging, molecular and atomic gas observations and simulations. As part of this survey 
 \cite{Young2010} (hereafter Paper IV) have presented an unbiased census of the molecular gas content of this sample of nearby early-types, and report that 22\% of optically-selected, morphologically-classified ETGs have substantial molecular gas reservoirs (10$^7$--10$^9$ M$_{\odot}$ of H$_2$). The brightest detections have been followed up with CO-interfeometry (Alatalo et al., in prep; see Section \ref{coobs}). Serra et al., in prep report a \hi\ detection rate of $\approx$40\% (for field galaxies with declination above 10$^{\circ}$ drawn from the same sample). These studies are the best available complete, unbiased surveys of the ISM in local ETGs, showing that a reasonable fraction of these systems have sizable gas reservoirs.

If galaxies form in a hierarchical manner, then these observations pose a challenge to the standard view that early-type galaxies join and remain on the red-sequence due to a lack of cold gas. One must either demonstrate that it is possible to create a tight red sequence without removing all of the cold ISM, or that galaxies can regenerate or acquire cold gas after joining the red sequence. 

The gas we detect in these galaxies could be a remnant, left over from the progenitors that formed the ETG.
Stellar population analyses (e.g. \citealt{Kuntschner:2006p1489}, \citealt{Kuntschner:2010p3471}, McDermid et al., in prep), however, show that the underlying stellar populations in these galaxies are old ($\gg1$\,Gyr). When combined with the fact that most ETGs show relaxed stellar kinematics \citep[][hearafter Paper III]{Emsellem2010} and evidence from deep imaging (Duc et al., submitted; Paper IX) this suggests that the last major merger that could have formed such systems was at least several gigayears ago. Unless the remnant gas can be made stable against star-formation this thus suggests that we are seeing regenerated or newly accreted gas.

Galaxies can regenerate a cold ISM through both \textit{internal} and \textit{external} processes. Stellar evolution models predict that an average stellar population returns on the order of half its stellar mass to the ISM over a Hubble time \citep{Jungwiert:2001p3209,Lia:2002p3210,Pozzetti:2007p3211,Martig:2009p2862}. This \textit{internal} gas return is dominated not by cataclysmic events such as supernovae, but by the long-term stellar mass loss from red giant branch, (post-)asymptotic giant branch stars and planetary nebulae \citep{Parriott:2008p2869,Bregman:2009p2870}.  The majority of the ejected mass from giant stars will shock heat and join the hot gas reservoir of the galaxy \citep{Parriott:2008p2869}.  It may then be possible for this material to cool, recombine, fall toward the galaxy centre and eventually become molecular, regenerating the cold gas reservoir. The fraction of gas which is able to cool from the hot halo in this way is however unknown. Mass loss from planetary nebulae may avoid being shock heated, and cool from the ionised phase directly without joining the galaxies hot halo \citep{Bregman:2009p2870}. Simulations have shown that this mass loss looses some angular momentum in this process, but generally it preserves the general sense of rotation of the parent star \citep[e.g.][]{Martig:2010p3475}. 

Stellar mass loss must be occurring in all ETGs at all times, at a rate depending on the number of stars present in each galaxy and its star formation history. One would thus expect many ETGs to have detectable molecular gas (assuming that some fraction of the hot gas reservoir does cool). 
Paper IV however found that only 22\% of ETGs have detectable molecular gas, and there is no dependence of the detection rate and molecular gas mass fraction on host luminosity. {It is of course possible that gas reservoirs exist in the other systems with a more quiescent recent star-formation history, resulting in low gas masses from stellar mass loss that are below our detection threshold.} The lack of correlation with the host galaxy properties, however, suggests that other mechanisms must be destroying or preventing the molecular gas from forming in 78\% of ETGs, and that stellar mass loss may not be the dominant method of acquiring cold gas.

Galaxies can also regain a cold ISM via \textit{external} processes, such as (major and minor) mergers and/or cold mode accretion from the intergalactic medium (IGM). In a minor merger or cold accretion some fraction of the 
gas can avoid being shocked and will fall to the centre of the galaxy and cool, often producing a central disc or rings \citep{Mazzuca:2006p3379,ElicheMoral:2009p2918}. In a major merger ejected gas can be re-accreted over the course of cosmic time. \hi\ in ETGs has long been thought to come almost exclusively from external sources \citep[e.g][]{Knapp:1985p3440}. More recently deep HI observations \citep[e.g.][]{Morganti:2006p1934} have confirmed that almost all \hi-detected field ETGs show signs of accretion, indicating an external origin for much of the atomic gas, with an average \hi\ accretion rate of 0.1 M$_{\odot}$ yr$^{-1}$ \citep{Oosterloo:2010p3376}. If the same is true for the molecular and ionised gas has yet to be determined.

Recent work has suggested one way to completely sidestep this problem, via `morphological quenching'  \citep{Martig:2009p2923}. In this scenario, red sequence galaxies are created not by removing all the cold ISM, but by making some fraction of it stable against gravitational collapse. The growth of a stellar spheroid via merging or secular processes increases the stellar density, and hence the epicyclic frequency and velocity dispersion in the centre of the galaxy. This can cause the Toomre stability criterion \citep{Toomre:1964p3272} to exceed unity, implying that the gas is stable against gravitational collapse. As the star formation rate would be very low, this stable cold ISM would persist for many billions of years, and hence would be detectable today in red galaxies. In some \atlas\ galaxies we do indeed see large amounts of \hi\ in a regular disk, but no corresponding young stellar population \citep{Morganti:2006p1934}. This process cannot be {how every ETG transits to the red sequence}, however, as such long-lived discs are not observed in all ETGs.

In summary, determining the dominant source of the cold ISM in ETGs is vital in order to understand their formation and evolution. If stellar mass loss can build up molecular reservoirs, then galaxies can transform themselves from spheroidal to disky systems over time and in isolation, and perhaps even (if morphological quenching does not intervene) evolve back into the blue cloud over cosmic timescales \citep[e.g.][]{Kannappan:2009p3436}. If however mergers are the dominant source of the gas, then star-formation episodes are likely to be short-lived and possibly violent, and the gas will not accumulate and regenerate.

Observationally the origin of the  gas may be addressed by comparing the angular momentum of the ISM with that of the underlying stellar population. Because of angular momentum conservation, stellar mass loss must produce gas that is kinematically aligned with the bulk of the stars which produced it. 
To first order, material from external sources can enter a galaxy with any angular momentum, so many such mismatches should be observable today, the exact fraction depending on the dynamical timescales for misaligned material to relax into the equatorial plane. Even in very old systems, one might thus expect to see an equal number of galaxies with counter and co-rotating gas if external accretion and mergers are important (and if there is no preferred accretion direction). 

A number of previous studies have used the misalignment of the angular momenta of gas and stars to constrain the origin of the ionized gas \citep[e.g.][]{Kannappan:2001p3451,Sarzi:2006p1474}, and the molecular gas \citep[e.g.][]{Young:2002p943,Young:2008p788,Crocker:2011p3291} in early-type galaxies. 
The goal of this paper is to use the \atlas\ complete volume limited sample of 260 ETGs to constrain the importance of externally acquired gas in this way.

In Section 2 of this paper we describe the 
observations our work is based upon, and present our method for extracting kinematic position angles. In Section 3 we describe the misalignment distributions obtained, and we discuss them further in Section 4. We conclude in Section 5 and discuss prospects for the future.

 \section{Observations and data analysis}
 To compare the orientation of the stellar and gaseous angular momenta, it is preferable to have full kinematic maps at comparable angular resolutions. \cite{Krajnovic:2008p1491} have shown that many ETGs have kinematically distinct stellar substructures, that may cause confusion in long-slit spectroscopic data. 
 Hence IFS data are desirable, where one retrieves full two-dimensional maps of the kinematics of the stars and ionised gas. These also allow classification of polar and misaligned structures, instead of confining alignment measures to purely co-rotating and counter-rotating bins (as one is forced to do with only long-slit spectroscopy). These 2D maps may be combined with mm-wave interferometry, to map the cold molecular component of the ISM at similar spatial resolutions. 
 
The galaxies we use in this work are drawn from the complete, volume-limited, morphologically selected \atlas\ sample (Paper I) of 260 ETGs brighter than $L>8.2\times10^9$ $L_{\odot,K}$, within $D<42$ Mpc. 
Paper I summarizes the main data-sets available for the survey galaxies. In this paper we combine the \atlas\ SAURON optical IFS stellar and ionised gas kinematic maps, \co\ kinematic maps, and \hi\ maps from the Westerbork Synthesis Radio Telescope (WSRT). The CO data are a combination of new observations from the Combined Array for Research in mm-wave Astronomy (CARMA) and literature data from the Plateau de Bure Interferometer (PdBI), Owens Valley Radio Observatory Interferometer (OVRO) and the Berkeley-Illinois-Maryland Array (BIMA). {These observations will be presented in full in Alatalo et al., in preparation, but we summarize them briefly below.}

\subsection{SAURON IFS data}
SAURON is an IFS built at Lyon Observatory 
and mounted at the Cassegrain focus of the 
William Herschel Telescope. It is based on the TIGER concept 
\citep{Bacon:1995p3377}, using a microlens array to sample the field of 
view. Details of the instrument can be found in \cite{Bacon:2001p1477}. All 
 galaxies were observed with the low-resolution mode 
of SAURON, covering a field of view of about 33\arcsec $\times$ 41\arcsec\
with 0$^{\prime\prime}_{.}$94 $\times$ 0$^{\prime\prime}_{.}$94 lenslets. Mosaicking was used 
to reach up to one effective radius (the radius encompassing half the light) when necessary. 

The SAURON observations for the \atlas\ galaxies, and the extraction of the stellar kinematics are described in detail in Paper I. In brief, for each target, individual datacubes were merged and analysed as described in \cite{Emsellem:2004p1497}, ensuring a minimum signal-to-noise ratio of 40 per spatial and spectral pixel 
using the binning scheme developed by \cite{Cappellari:2003p3284}. 
The SAURON stellar kinematics were derived using a penalized 
pixel fitting routine \citep{Cappellari:2004p3283}, providing
parametric estimates of the line-of-sight velocity distribution for each spaxel. During the extraction of the stellar kinematics, the GANDALF code \citep{Sarzi:2006p1474} was used to simultaneously extract the ionised gas kinematics.  
The mean stellar velocity maps for the \atlas\ galaxies are presented in \cite{Kraj2010}, hereafter Paper II. 

\subsection{CARMA data} 
\label{coobs}

As part of the \atlas\ survey, all galaxies that were detected in \co(1-0) {(from single pointings of the IRAM-30m telescope at the galaxy centre; see Paper IV)} with an integrated flux greater than 19 Jy \kms\ that do not have interferometric data already available in the literature have been observed with CARMA \citep{Bock:2006p2806}. Full details of this interferometric survey can be found in Alatalo et al. (in preparation), but we summarize the observations here. 
 
Observations of the detected sample galaxies were taken between 2008 and 2010, mainly in the D-array configuration, providing a spatial resolution of 4-5\arcsec.
\co(1-0) was observed using narrow-band correlator configurations, providing at least 3 raw channels per 10 \kms\ binned channel whilst ensuring adequate velocity coverage for all galaxies.

Bright quasars were used to calibrate the antenna-based gains and for passband calibration. The data were calibrated and imaged using the Multichannel Image Reconstruction, Image Analysis and Display (MIRIAD) software package \citep{Sault:1995p2768}. Total fluxes {and velocity widths} were compared with the IRAM 30m single-dish observations to ensure that large proportions of the fluxes were not being resolved out. 

A total of 27 galaxies included in this work have been observed and all were detected at greater than 3$\sigma$ with CARMA (IC\,676, IC\,719, IC\,1024, NGC\,1222, NGC\,2697, NGC\,2764, NGC\,2824, NGC\,3182, NGC\,3607, NGC\,3619, NGC\,3626, NGC\,3665, NGC\,4119, NGC\,4324, NGC\,4429, NGC\,4435, NGC\,4694, NGC\,4710, NGC\,4753, NGC\,5379, NGC\,5866, NGC\,6014, NGC\,7465, PGC\, 029321, PGC\,058114, UGC\,06176 and UGC\,09519). 
We also include here galaxies for which CO(1-0) maps are already available from the literature, mostly from SAURON survey \citep{deZeeuw:2002p1496} follow-ups. 
These are NGC\,0524, NGC\,2685, NGC\,2768, NGC\,3032, NGC\,3489, NGC\,4150, NGC\,4459, NGC\,4476, NGC\,4477, NGC\,4526 and NGC\,4550. Full references for the data used in this work are listed in Table \ref{datatable}. This makes for a total of 38 sample galaxies that have CO(1-0) interferometric data.

This sample of mapped galaxies are the brightest $\approx$2/3 of the CO detections from Paper IV. The mapped galaxies are thus biased towards higher molecular gas masses, but are statistically indistinguishable (with a Mann--Whitney U (MW-U) or Kolmogorov-Smirnov test) from the full sample of CO detected systems in terms of their molecular mass fractions, local environment, and host galaxy properties. The mapped galaxies also fall on the same CO Tully-Fisher relation \citep[][hearafter Paper V]{Davis:2011p3472}. Other than molecular gas mass, the only variable in which our sample may have some bias is in cluster membership. Due to the fact that the Virgo cluster is relatively nearby (and hence the flux from a given amount of molecular gas is higher) we have mapped a greater percentage of cluster members than field galaxies. We do not however expect this to affect our conclusions. Overall we consider that the CO sample discussed from this point should be reasonably free of biases, however it is worth remembering that we do not full sample the parameter space at smaller molecular gas masses.

In this work we make use of the CO velocity fields which will be presented in Alatalo et al, in prep. These mean velocity maps 
were produced by the masking method: each fully calibrated and cleaned image 
cube was smoothed along both the spatial and velocity axes, and the 
smoothed cube was clipped at 
$\approx$2.5 times the rms noise in a 
channel. The clipped version of the smoothed cube was then used 
as a mask to define a three-dimensional volume in the original, 
unsmoothed cleaned cube, over which we calculate the first velocity moment \citep[see][]{Regan:2001p3275,Young:2008p788}. An identical method was used in \cite{Young:2008p788,Crocker:2008p946,Crocker:2009p3262} and \cite{Crocker:2011p3291}, and we refer readers to these works for examples of the data quality achieved.

\subsection{Measuring kinematic position angles}

The apparent (projected) direction of the angular momentum vector of any tracer can be found by estimating its kinematic position angle $\phi$, which lies normal to the angular momentum vector. The kinematic position angle (PA) is defined as the counter-clockwise angle between north and a line which bisects the tracer's velocity field, measured on the receding side.  

One can then define the kinematic misalignment angle $\psi$ as the difference between the kinematic PA of two galactic components. In this paper, we will consider misalignments between the molecular gas, warm ionised gas and stellar angular momentum:
\begin{eqnarray}
&& \psi_{\rm mol-star} \equiv |\phi_{\rm molecular\,gas}- \phi_{\rm star}|,\\
 && \psi_{\rm mol-ion} \equiv |\phi_{\rm molecular\,gas}- \phi_{\rm ionised\,gas}|,\\
 && \psi_{\rm ion-star} \equiv |\phi_{\rm ionised\,gas} - \phi_{\rm star}|,
\end{eqnarray}

These quantities are defined to lie in the range 0-180$^{\circ}$.

In this work, the kinematic PA of the stars is taken from Paper II (Table 1, column 4 in Appendix C, and available online at http://www.purl.org/atlas3d). The molecular and ionised gas kinematic PAs are presented here for the first time, and are measured using the same method (the \textsc{fit\_kinematic\_pa} routine\footnote{Available from http://purl.org/cappellari/idl} described in Appendix C of \citealt{Krajnovic:2006p2929}).

\begin{figure*}
\subfigure{\includegraphics[height=3.8cm,angle=0,clip,trim=0cm 0cm 0cm 0.5cm]{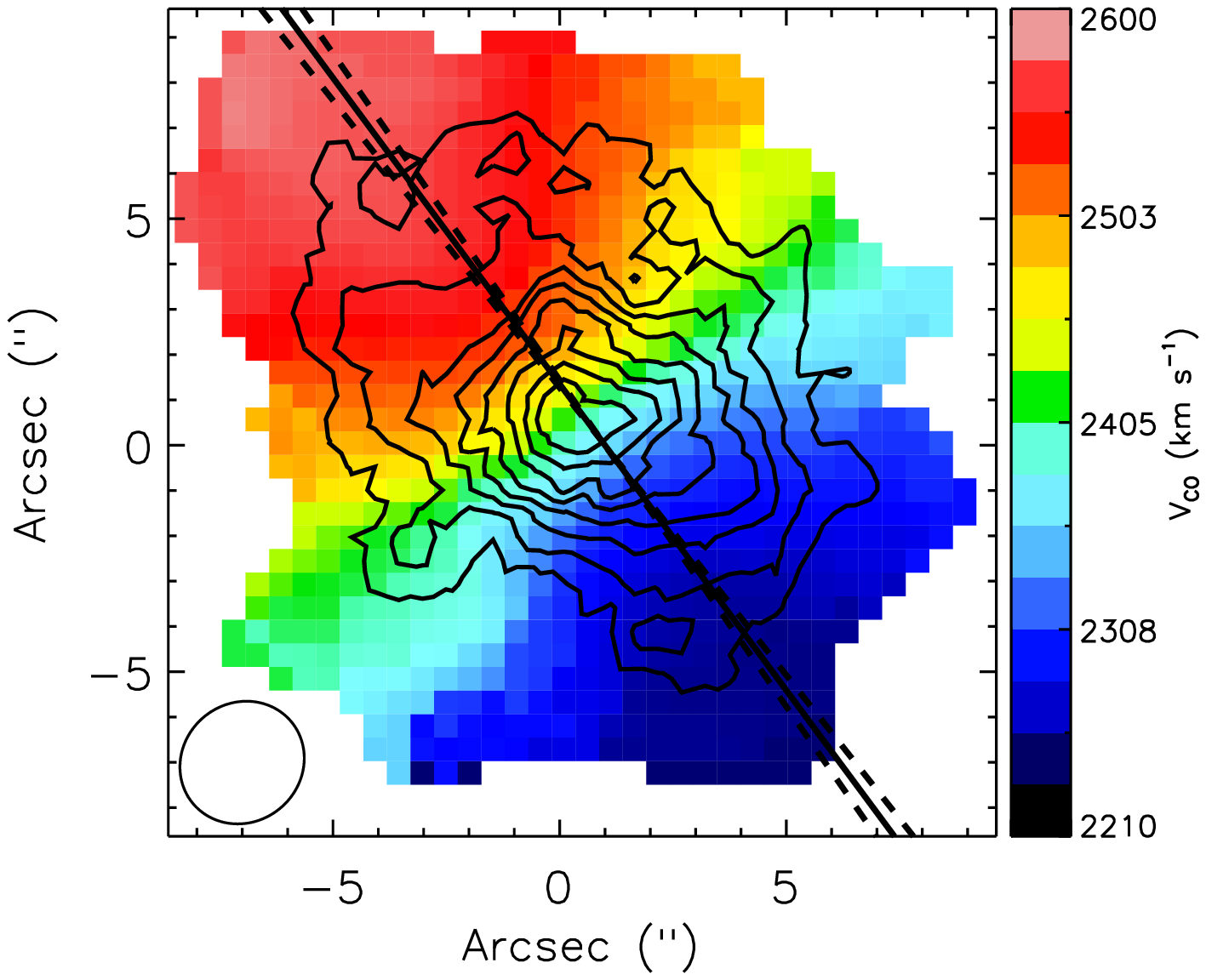}}
\subfigure{\includegraphics[height=3.8cm,angle=0,clip,trim=0cm 0cm 0cm 1.0cm]{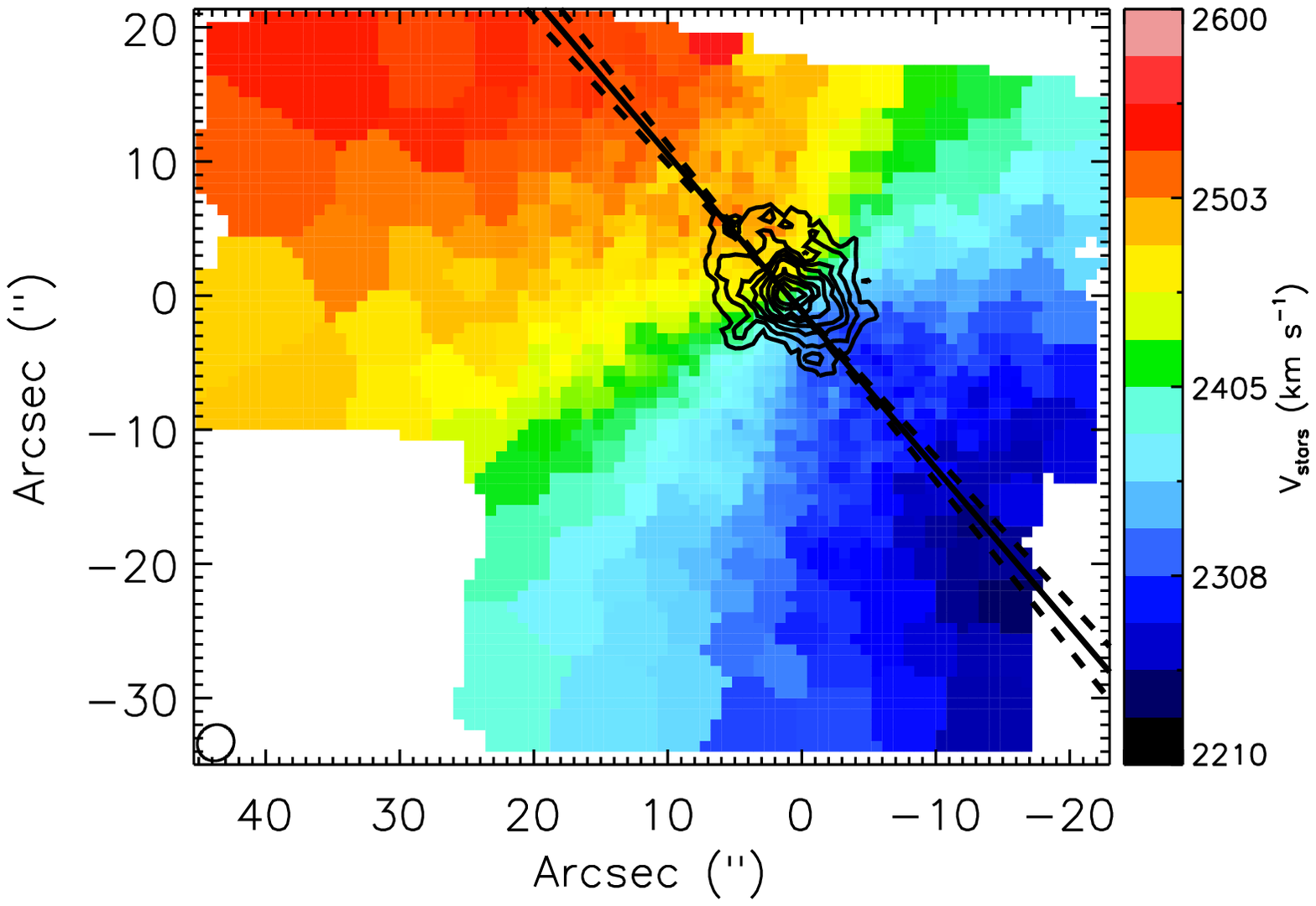}}
\subfigure{\includegraphics[height=3.8cm,angle=0,clip,trim=0cm 0cm 0cm 0.5cm]{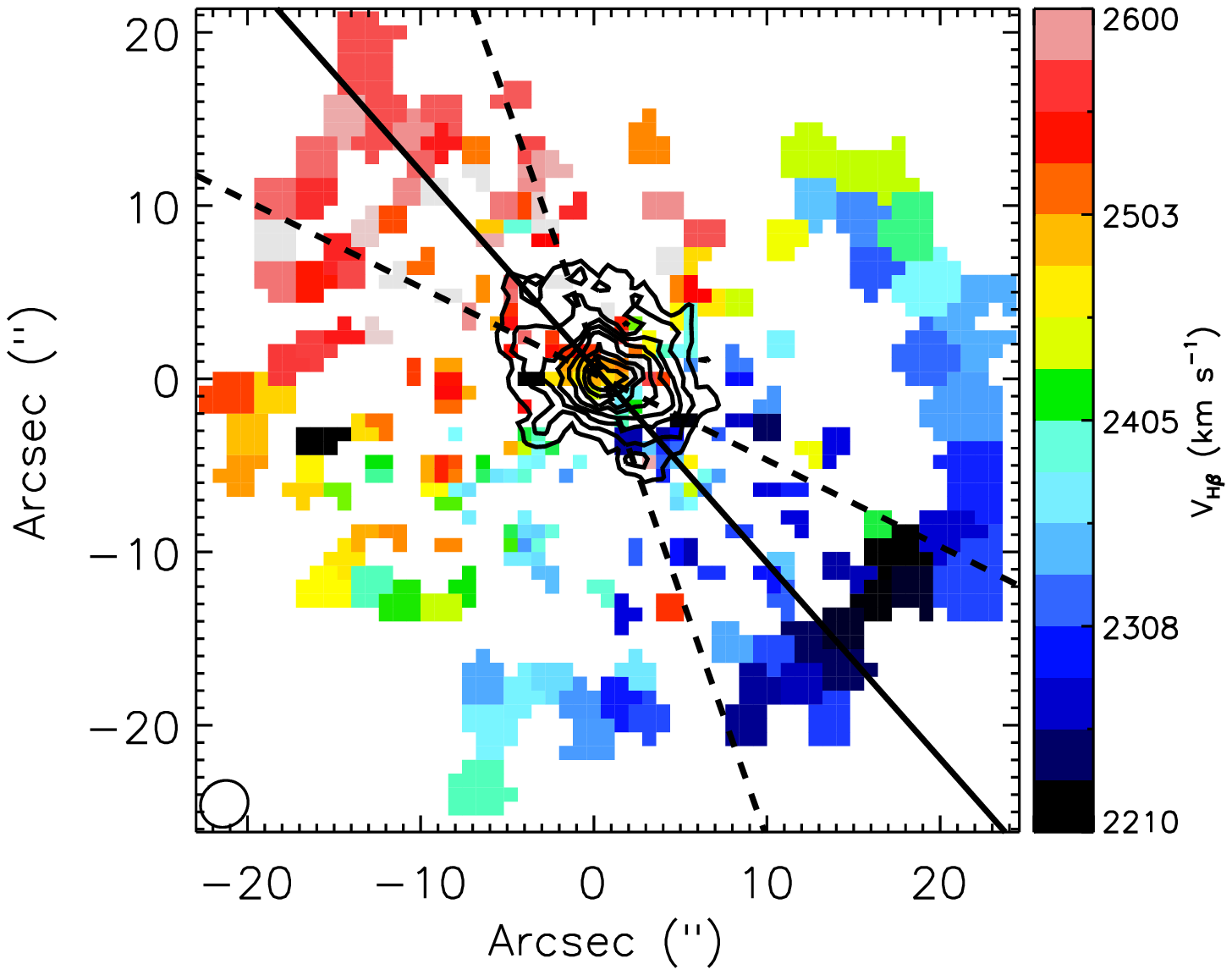}}\\
\subfigure{\includegraphics[height=3.8cm,angle=0,clip,trim=0cm 0cm 0cm 0.5cm]{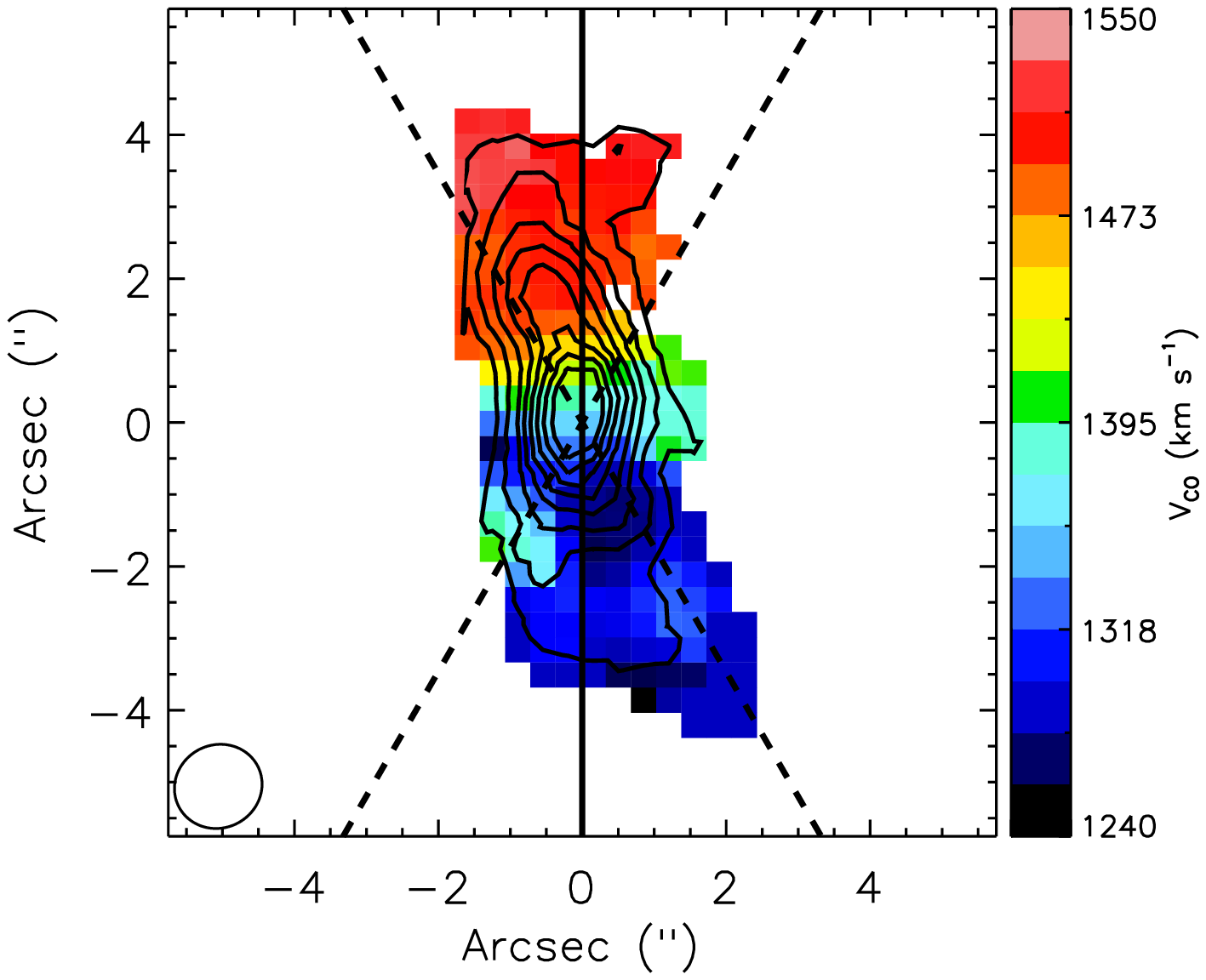}}
\subfigure{\includegraphics[height=3.8cm,angle=0,clip,trim=0cm 0cm 0cm 0.5cm]{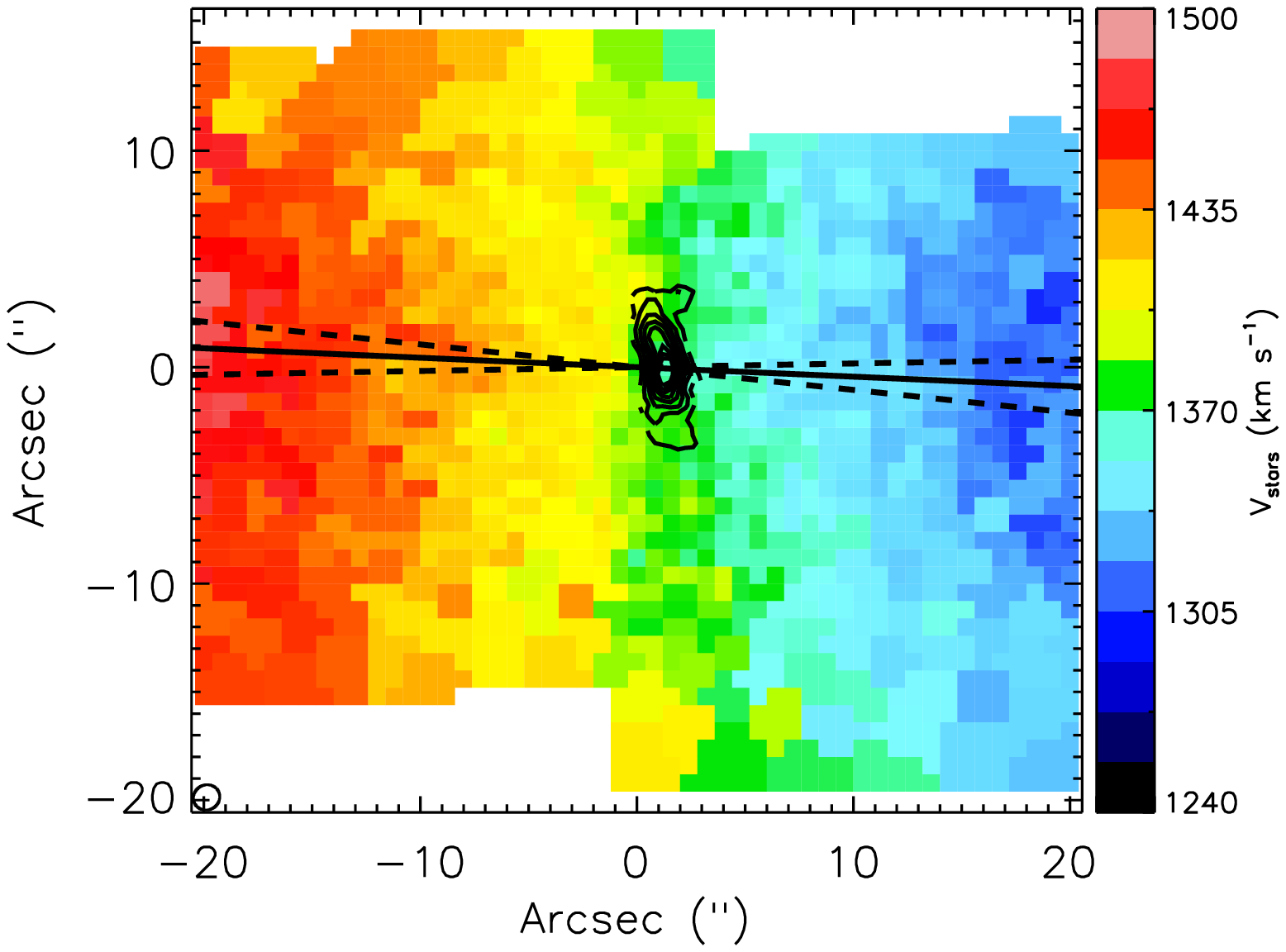}}
\subfigure{\includegraphics[height=3.8cm,angle=0,clip,trim=0cm 0cm 0cm 0.5cm]{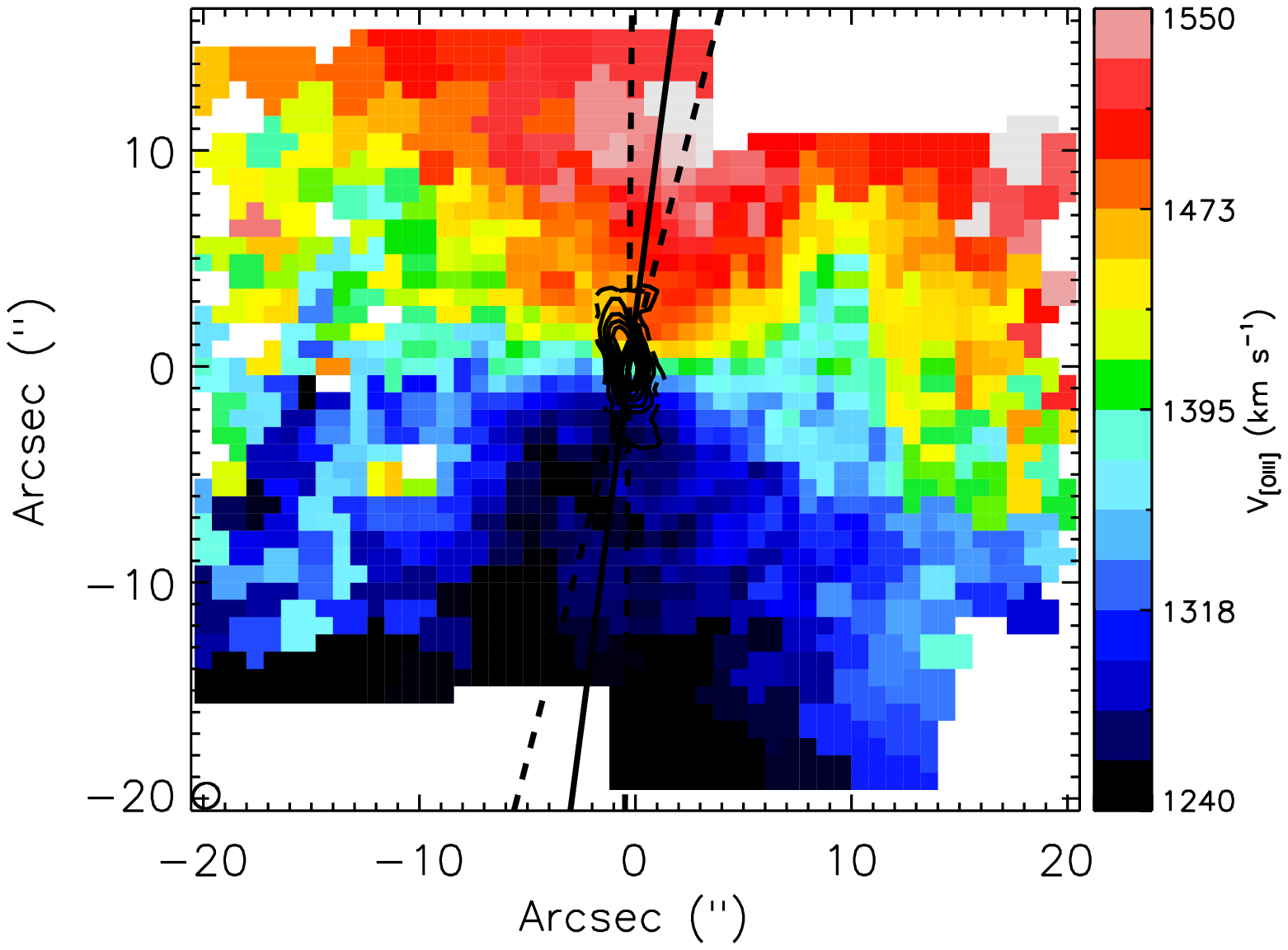}}
\caption{\small Examples of the data for one of the galaxies with the best constrained molecular gas PA (NGC\,0524, top), and the galaxy with the worst constrained molecular gas PA (NGC2768, bottom). \textit{Left:} CO velocity field (colours). \textit{Middle:} Stellar velocity fields from the SAURON IFU \citep[data from][]{Emsellem:2004p1497}. \textit{Right:} H$\beta$ emission line kinematics from the SAURON IFU \citep[data from][]{Sarzi:2006p1474}. These are overlaid with the CO integrated intensity (black contours) and the best fitting molecular gas/stellar kinematic PA, and its associated error as calculated in this paper (solid black line, and dashed black lines respectively). The CO moments are taken from \citealt{Crocker:2008p946} (NGC2768) and \citealt{Crocker:2011p3291} (NGC0524).  The SAURON IFU coverage of NGC2768 extends beyond the range of the plots shown above. 
}
\label{kinpaexample}
\end{figure*}

In this paper we were able to measure molecular gas kinematic position angles for all the galaxies with mapped molecular gas. The values of the kinematic PA for these galaxies are listed in Table \ref{datatable}. We give two examples of CO velocity fields overlaid with their derived kinematic PA and the associated error in Figure \ref{kinpaexample}.  Ionised gas kinematic position angles were measured only on galaxies where the emission appeared coherent, with a regular velocity field, and was not very confined to the very central regions (as this is indicative of template mismatch within the GANDALF package; see \citealt{Sarzi:2006p1474}).  These criteria were satisfied in 132 of the 260 sample galaxies (51\%). Where the ionised gas is only detected over a small region, we have re-run the  \textsc{fit\_kinematic\_pa} code using only this region to reduce the effect of low signal to noise ratio bins on our PAs determinations. Table \ref{datatable} contains the ionised gas kinematic PA for all CO mapped galaxies in this work. Tables \ref{datatable1} and and \ref{srtable} contain the kinematic misalignment between the ionised gas and stellar components for all fast-rotating galaxies and slow rotating galaxies respectively, which were not mapped in CO, but which have a significant enough ionised gas detection to allow a kinematic PA measurement.

\hi\ data was available for 29 fast-rotating galaxies that overlap with the CO and ionised gas sample considered in this paper. The large difference in angular scale probed make comparisons between the extended \hi\ and centrally concentrated molecular and ionized gas challenging, but potentially powerful. We do not extract kinematic position angles for the \hi, {but discuss the general trends present from visual inspection of the \hi\ velocity fields} in Section \ref{hi}. More detailed analysis of the \hi\ velocity fields and intensity maps will take place in a future paper in this series.

In real galaxies, there can be kinematic substructures (e.g. bars, kinematically decoupled cores) that may or may not share the same kinematic PA. In these cases, the \textit{fit\_kinematic\_pa} routine returns the value applicable to the bulk of the material (as it is area, rather than luminosity weighted). Such substructures can, however, skew the measured kinematic PA, in some cases by tens of degrees. When a bar is present, Paper II has shown that the global kinematic PA of the stars is uncertain with a standard deviation of $\approx$10$^{\circ}$. In general however the stellar kinematic position angle is a good estimate of the PA of the line of nodes even when a bar exists. Bars would be expected to have a stronger influence on the gaseous components than on the stars. 

It is worth remembering that we always compare the kinematic PAs of the molecular gas to that of the main body of the stars. Some stellar velocity fields show a  kinematically decoupled core (KDC), a stellar subcomponent with misaligned rotation from the rest of the stellar body. These KDCs generally extend only over a small central region however, and hence do not significantly affect our results.

The mm interferometric data have a poorer angular resolution than the optical data ($\approx$4-5\arcsec\ vs 1-2\arcsec). In order to test how this affects the determination of the kinematic PA, we performed a Monte-Carlo simulation. A synthetic velocity field with a known PA was created using an empirical galaxy rotation curve from \cite{Roscoe:1999p3276}. We then added Gaussian noise to simulate the average signal to noise ratio of our data. We varied the inclination, and the angular extent of the emitting material while keeping a fixed beam size (which is identical to changing the angular resolution) using a fixed spatial sampling of 3 pixels per beam, and measured the kinematic PA using the \textit{fit\_kinematic\_pa} routine. Changing the inclination from edge-on to face-on had little effect of the determination on the PA, apart from very face on inclinations where the velocity range becomes comparable to the channel width. This happens only at inclinations less than 2.8 degrees for a typical galaxy with a maximum velocity of 200 \kms, and 10 \kms\ channel width. Varying the source size from 50 to 2.5 beams across  (or equivalently 1\arcsec\ to 20\arcsec\ angular resolution) caused a scatter of $\approx$10$^{\circ}$ in the determined kinematic PA, with no systematic offset. The true PA was always within the 3$\sigma$ error bars returned by the routine. 

{We also used this simulation to test the effect of signal to noise on our PA determinations. The average RMS deviation from the bi-(anti)symmetric velocity map created by the \textit{fit\_kinematic\_pa} is $\approx$15 \kms\ for the real CARMA data used in this work (with a maximum of 35 \kms). Varying the simulated RMS noise from 5 \kms\ to 50 \kms\ scattered the derived kinematic PAs by $\approx$10$^{\circ}$. Once again this induced no systematic offset, and the true PA was always within the 3$\sigma$ error bars returned by the routine.}

The largest error on the observed molecular gas PA (as estimated by the \textit{fit\_kinematic\_pa} routine) is 30$^{\circ}$, and the average (1$\sigma$) error on the molecular gas PAs is 7$^{\circ}$, consistent with the predicted error from our simulation.  
In this paper we choose to consider kinematic misalignments between the gas and stars of $>30^{\circ}$ as being significantly misaligned. This conservative estimate ensures that we can set a robust lower limit on the importance of externally accreted material.

\begin{table*}
\caption{Kinematic parameters of the \atlas\ early-type CO-rich galaxies used in this paper.}
\begin{tabular*}{1\textwidth}{@{\extracolsep{\fill}}r r r r r r r r r r r r r r r l}
\hline
Name & $\phi_{\rm mol}$ & $\Delta\phi_{\rm mol}$ & $\phi_{\rm ion}$ & $\Delta\phi_{\rm ion}$ & $\psi_{\rm mol-star}$ & $\Delta\psi_{\rm mol-star}$ & $\psi_{\rm mol-ion}$ & $\Delta\psi_{\rm mol-ion}$ & $\psi_{\rm ion-star}$ & $\Delta\psi_{\rm ion-star}$ & Ref. \\
& (deg) & (deg) & (deg) & (deg) & (deg) & (deg) & (deg) & (deg) & (deg) & (deg) & \\
(1) & (2) & (3) & (4) & (5) & (6) & (7) & (8) &(9)&(10)&(11) & (12)\\
\hline
IC0676 & 16.5 & 10.0 & 348.5 & 26.9 & 2.0 & 15.8 & 28.5 & 29.6 & 30.0 & 29.6 & 1\\
IC0719 & 229.0 & 3.5 & 232.5 & 2.5 & 177.0 & 26.2 & 3.5 & 26.1 & 173.5 & 26.1 & 1\\
IC1024 & 24.5 & 12.8 & 31.5 & 8.2 & 5.0 & 16.7 & 7.0 & 13.6 & 2.0 & 13.6 & 1\\
NGC0524 & 36.5 & 1.8 & 41.5 & 21.9 & 4.0 & 2.7 & 5.0 & 22.0 & 1.0 & 22.0 & 6\\
NGC1222 & 33.0 & 1.5 & 54.5 & 21.5 & 10.0 & 9.4 & 21.5 & 23.4 & 11.5 & 23.4 & 1\\
NGC2685 & 105.0 & 10.0 & 109.5 & 13.8 & 68.5 & 10.3 & 4.5 & 14.0 & 73.0 & 14.0 & 2\\
NGC2697 & 301.5 & 1.8 & 300.0 & 3.1 & 0.0 & 3.6 & 1.5 & 4.4 & 1.5 & 4.4 & 1\\
NGC2764 & 202.5 & 1.8 & 189.5 & 5.2 & 6.5 & 7.0 & 13.0 & 8.6 & 6.5 & 8.6 & 1\\
NGC2768 & 180.0 & 30.0 & 187.5 & 6.1 & 87.5 & 30.2 & 7.5 & 7.0 & 95.0 & 7.0 & 4\\
NGC2824 & 161.5 & 1.8 & 159.5 & 7.5 & 2.0 & 3.3 & 2.0 & 8.0 & 0.0 & 8.0 & 1\\
NGC3032 & 92.5 & 10.0 & 62.5 & 19.1 & 179.0 & 14.9 & 30.0 & 22.0 & 151.0 & 22.0 & 3\\
NGC3182 & 331.5 & 7.2 & 316.5 & 6.0 & 11.0 & 10.0 & 15.0 & 9.2 & 4.0 & 9.2 & 1\\
NGC3489 & 67.5 & 1.8 & 78.5 & 5.3 & 5.0 & 3.3 & 11.0 & 6.0 & 6.0 & 6.0 & 6\\
NGC3607 & 302.5 & 2.5 & 302.5 & 0.5 & 1.0 & 2.9 & 0.0 & 1.6 & 1.0 & 1.6 & 1\\
NGC3619 & 74.5 & 4.2 & 76.0 & 0.5 & 22.5 & 5.2 & 1.5 & 3.0 & 24.0 & 3.0 & 1\\
NGC3626 & 169.5 & 1.8 & 165.0 & 3.2 & 170.0 & 1.9 & 4.5 & 3.3 & 174.5 & 3.3 & 1\\
NGC3665 & 219.5 & 2.0 & 209.5 & 2.2 & 14.0 & 2.8 & 10.0 & 3.0 & 4.0 & 3.0 & 1\\
NGC4119 & 296.0 & 12.0 & 292.5 & 18.6 & 4.5 & 13.6 & 3.5 & 19.7 & 1.0 & 19.7 & 1\\
NGC4150 & 146.0 & 10.0 & 168.5 & 10.1 & 1.5 & 11.9 & 22.5 & 12.0 & 21.0 & 12.0 & 3\\
NGC4324 & 232.0 & 1.8 & 239.0 & 6.8 & 6.0 & 5.5 & 7.0 & 8.6 & 1.0 & 8.6 & 1\\
NGC4429 & 82.0 & 2.0 & 92.5 & 2.0 & 4.5 & 3.2 & 10.5 & 3.2 & 6.0 & 3.2 & 1\\
NGC4435 & 201.0 & 1.8 & 198.6 & 3.1 & 8.5 & 2.5 & 2.4 & 3.6 & 6.1 & 3.6 & 1\\
NGC4459 & 269.0 & 5.0 & 281.5 & 1.5 & 11.5 & 5.6 & 12.5 & 2.9 & 1.0 & 2.9 & 3\\
NGC4476 & 208.0 & 2.5 & 219.0 & 7.5 & 1.5 & 11.8 & 11.0 & 13.7 & 12.5 & 13.7 & 1\\
NGC4477 & 227.0 & 3.0 & 226.5 & 1.6 & 25.5 & 6.0 & 0.5 & 5.5 & 26.0 & 5.5 & 6\\
NGC4526 & 288.0 & 3.0 & 294.5 & 4.7 & 0.5 & 3.5 & 6.5 & 5.0 & 6.0 & 5.0 & 3\\
NGC4550 & 355.0 & 10.0 & 358.5 & 9.7 & 3.5 & 10.3 & 3.5 & 10.0 & 0.0 & 10.0 & 5\\
NGC4694 & 155.5 & 27.5 & 167.0 & 45.0 & 169.0 & 33.6 & 11.5 & 48.9 & 157.5 & 48.9 & 1\\
NGC4710 & 207.0 & 10.0 & 208.5 & 2.2 & 0.5 & 10.7 & 1.5 & 4.4 & 1.0 & 4.4 & 1\\
NGC4753 & 93.0 & 2.8 & 85.5 & 1.5 & 4.5 & 3.8 & 7.5 & 2.9 & 3.0 & 2.9 & 1\\
NGC5379 & 66.0 & 20.0 & 77.0 & 5.0 & 5.0 & 22.4 & 11.0 & 11.2 & 16.0 & 11.2 & 1\\
NGC5866 & 127.0 & 2.5 & 123.0 & 4.5 & 2.0 & 3.9 & 4.0 & 5.1 & 2.0 & 5.4 & 1\\
NGC6014 & 139.5 & 6.2 & 151.5 & 10.2 & 7.5 & 10.5 & 12.0 & 13.3 & 4.5 & 13.3 & 1\\
NGC7465 & 106.0 & 3.8 & 109.5 & 9.5 & 60.5 & 29.2 & 3.5 & 30.5 & 57.0 & 30.5 & 1\\
PGC029321 & 76.0 & 27.5 & 105.5 & 27.2 & 19.5 & 46.1 & 29.5 & 46.0 & 49.0 & 46.0 & 1\\
PGC058114 & 94.5 & 7.0 & 93.1 & 3.1 & 152.5 & 12.2 & 1.4 & 10.5 & 153.9 & 10.5 & 1\\
UGC06176 & 201.0 & 8.5 & 214.0 & 8.5 & 0.5 & 30.2 & 13.0 & 30.2 & 13.5 & 30.2 & 1\\
UGC09519 & 177.5 & 5.5 & 172.0 & 4.5 & 72.0 & 8.0 & 5.5 & 7.3 & 77.5 & 7.3 & 1\\
\hline
\end{tabular*}
\parbox[t]{1 \textwidth}{ \textit{Notes:} Columns 2-5 list the kinematic PAs for the molecular gas and ionised gas (even columns), and their respective errors (odd columns). Columns 6-11 list the misalignments between the molecular gas, ionised gas and stars (even columns) and and their respective errors (odd columns). The errors quoted on the kinematic PAs are those calculated by the \textsc{fit\_kinematic\_pa} routine. The kinematic PA of the main body of the stars for each galaxy is taken directly from Paper II (Table 1 in Appendix C, Column 4; available online at http://www.purl.org/atlas3d). Column 12 lists the data reference: (1) Alatalo et al., in prep (2) \cite{Schinnerer:2002p981}, (3) \cite{Young:2008p788}, (4) \cite{Crocker:2008p946}, (5) \cite{Crocker:2009p3262}, (6) \cite{Crocker:2011p3291}.}

\label{datatable}

\end{table*}

\section{Results}
\subsection{Molecular gas and stars}
\label{molstar}
\begin{figure}
\includegraphics[scale=0.5,angle=0,clip,trim=0.6cm 0cm 0.7cm 0.7cm]{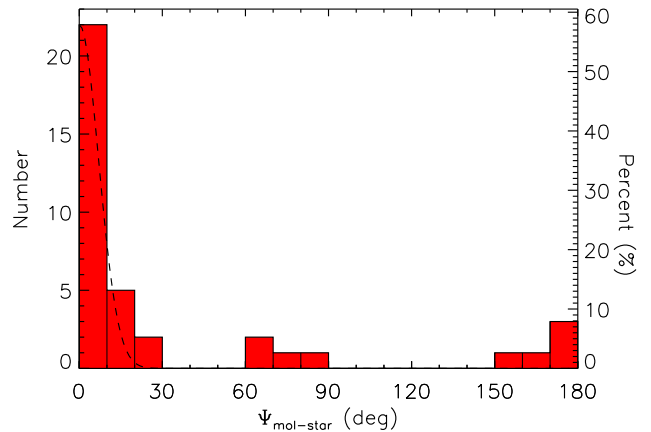}
\caption{\small Histogram showing the kinematic misalignment angle between the molecular gas and the stars for all of the galaxies listed in Table \ref{datatable}. The dashed line overplotted is a normalized Gaussian distribution with its center at zero and a standard deviation of 7$^{\circ}$, showing the expected scatter.}
\label{mol_star}
\end{figure}

A direct comparison between the kinematic PAs of the molecular gas and the main body of the stars reveals two populations (Figure \ref{mol_star}). We find 24$\pm$7\% (9 out of 38 galaxies) have their molecular gas misaligned ($\Psi > 30^{\circ}$) with respect to the stars, setting a lower limit on the number of galaxies where the gas is likely to have an external origin. The remaining 76$\pm$7\% (29 out of 38 galaxies) have their molecular gas aligned with the stars ($\Psi < 30^{\circ}$) with the stars, consistent with an internal origin. 
The uncertainties quoted above include only the formal binomial counting errors.

Several individual galaxies have a KDC coincident with the molecular gas. NGC\,4476 is a pathological example, as it has a large stellar KDC, kinematically aligned with (and likely formed from) the molecular gas, while the stars have little sense of rotation outside this region within the SAURON IFS field of view. PPAK IFS stellar kinematics suggest that the stellar velocity field may become inverted outside the SAURON field of view (Crocker et al., in prep), hence the molecular gas could be considered to be counter-rotating with respect to the stellar body at large radii. For consistency, however, we use only the SAURON stellar kinematics in this work (probing roughly one effective radius), and hence label NGC\,4476's molecular gas as co-rotating. It is worth bearing in mind, however, that the stellar kinematics in the outer parts of all of these galaxies could be different from the kinematics in the central regions. 
Discounting the few galaxies such as this, where the origin of the gas is hard to judge would not significantly alter our conclusions.

\subsection{Molecular and ionised gas}
\label{iongas}

\begin{figure}
\includegraphics[scale=0.5,angle=0,clip,trim=0.6cm 0cm 0.7cm 0.7cm]{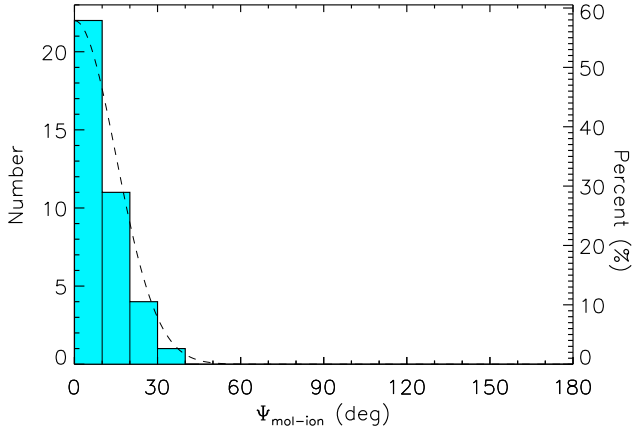}
\caption{\small Histogram showing the kinematic misalignment angle between the molecular gas and the ionised gas for all the galaxies listed in Table \ref{datatable}. The dashed line overplotted is a normalized Gaussian distribution with its center at zero and a standard deviation of 15$^{\circ}$, showing the expected scatter. }
\label{mol_ion}
\end{figure}

The analysis presented in Section \ref{molstar} can be repeated, comparing the apparent angular momenta of the molecular and ionised gas. This results in Figure \ref{mol_ion}. The molecular gas has a kinematic misalignment angle with the ionised component of less than 40$^{\circ}$ in every galaxy (and generally much better than this), and within the errors quoted all galaxies are consistent with having a kinematic misalignment angle of less than 10.5$^{\circ}$. The overall distribution is consistent with a Gaussian error distribution with centre at zero and a standard deviation of 15$^{\circ}$, as shown in Figure \ref{mol_ion}. 

The galaxies which have a misalignement between 15$^{\circ}$ and 40$^{\circ}$ are IC\,0676, NGC\,1222, NGC\,3032, NGC\,4150 and PGC\,029321. These small misalignments can always be understood as arising from small-scale kinematic substructures within the galaxies affecting the ionised component, that are usually unresolved at the resolution of our CO interferometry. 
IC\,0676 has its ionised gas kinematic PA affected by a bar. NGC\,1222 is a highly disturbed star-bursting system, that may be in the process of undergoing a three-way merger \citep{Beck:2007p2941}.  As discussed above, NGC\,3032 and NGC\,4150 have KDCs that may be affecting the gas properties \citep{McDermid:2006p1493, Sarzi:2006p1474}. The ionised and molecular material in PGC\,029321 has a small angular extent, and as we showed above this can increase the error in the kinematic PA determination.

The uncertainty on ionised gas PA measurements is larger than the average 10$^{\circ}$ we have estimated for the molecular gas PAs. As discussed above, this is because the ionised gas is often patchy, has a smaller angular extent than the molecular gas, and can be less relaxed. Despite this increased uncertainty, and greater intrinsic variance due to small substructures in the galaxies, our conservative threshold for misalignment ($\Psi>30^{\circ}$) ensures that we still set a strong lower limit on the importance of externally accreted material. Despite the increased uncertainty, we only show plots of the ionised gas misalignment from here on, taking advantage of the increased number of galaxies available to improve our statistics. When one examines the molecular gas kinematic misalignments all results remain unchanged.

{One galaxy (NGC\,3032) has a molecular to ionized gas kinematic misalignment $\gtsimeq$30$^{\circ}$ cut-off we have taken to denote the boundary between aligned and misaligned gas. As the distribution of misalignments closely follows the gaussian error distribution however, we suggest that this is due to observational scatter. }
We therefore conclude that the ionised and molecular material are always kinematically aligned with each other, even when they are kinematically misaligned with respect to the stars, strongly suggesting they share a common origin  \citep[as previously discussed in a subsample of these galaxies by][]{Crocker:2011p3291}. This is discussed further in Section \ref{ion_discuss}. 

{Assuming 30$^{\circ}$ rather than 40$^{\circ}$ for the cut-off between aligned and misaligned gas can at most bias our results by $\approx$4\% (this is the area under the error distribution at misalignments greater than 30$^{\circ}$, normalized by the total area under this error distribution curve.) For the typical number of galaxies included in our analysis this can result in the misclassification of 1-2 galaxies.}

\subsection{Ionised gas and stars}

As the molecular and ionised gas are always aligned and are likely to share a common origin, investigating the kinematic misalignment between the ionised gas and the stars can therefore also shed light on the origin of both of these gas phases in ETGs. 

The \atlas\ ionised gas detection rate (above an integrated equivalent width of 0.02 \AA\ in either H$\beta$ or [OIII]) is $\approx$73$\pm$3\% in the field, and $\approx$47$\pm$6\% in the Virgo cluster. These figures are comparable with the previous SAURON survey ionised gas detection rate of 86$\pm$6\% in the field and 55$\pm$12\% in the Virgo cluster \citep{Sarzi:2006p1474}. We are hence able to repeat the analysis in Section \ref{molstar}, 
with the benefit of increased number statistics. 

Table \ref{datatable1} contains the kinematic misalignment between the ionised gas and stars for all fast-rotating galaxies with a significant enough ionised gas detection to allow a kinematic PA measurement. 

\subsubsection{Slow-rotators}

Slow-rotators as a class are round, massive, dispersion dominated, mildly triaxial ETGs that generally have old stellar populations, and large KDCs (Paper III).  Ionised gas is detected in slow-rotators \citep{Sarzi:2006p1474}, while molecular gas almost never is (Paper IV). 
Ionised gas is hence the best tracer to help us understand the origin of the gas in the central parts of these systems. The stellar kinematic PA can be poorly defined in a slow-rotator, which have little or no coherent rotation by definition, and this translates to large uncertainties in the misalignments. Despite this it is possible to estimate the kinematic misalignment for 22 of the 36 slow rotating galaxies in the \atlas\ sample. The derived misalignments are listed in Table \ref{srtable}.

\begin{figure}
\includegraphics[scale=0.5,angle=0,clip,trim=0.6cm 0cm 0.7cm 0.7cm]{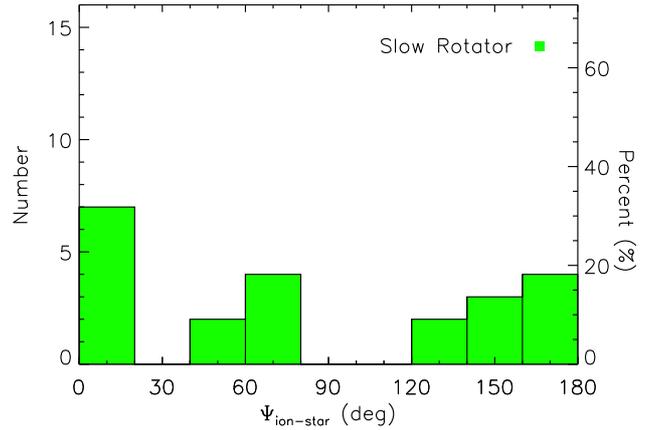}
\caption{\small Histogram showing the kinematic misalignment angle between the ionised gas and the stars for all the slow-rotating galaxies with measurable kinematic misalignments. These systems are listed in Table \ref{srtable}.}
\label{srion_star}
\end{figure}

When slow rotators are considered by themselves (Figure \ref{srion_star}), they display a flat distribution of misalignments between the ionised gas and the stars, suggesting the dominant source of (ionised) gas is external \citep[as was seen previously with fewer galaxies by][]{Sarzi:2007p3289}. A MW-U test finds no statistically significant evidence that the observed kinematic misalignments (in Figure \ref{srion_star}) are not drawn from a uniform underlying parent distribution.

\subsubsection{Fast-rotators}

\begin{figure}
\includegraphics[scale=0.5,angle=0,clip,trim=0.6cm 0cm 0.7cm 0.7cm]{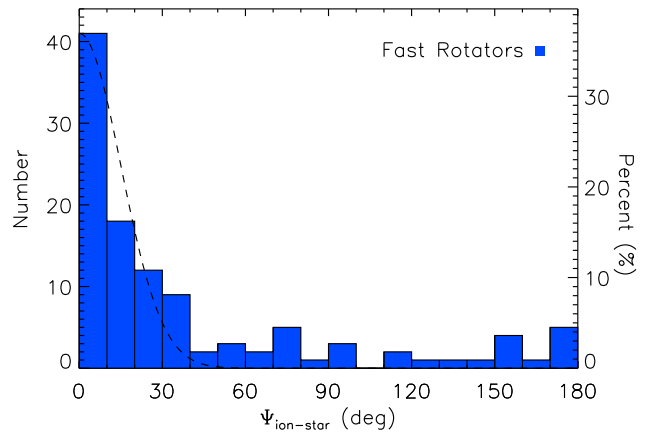}
\caption{\small Histogram showing the kinematic misalignment angle between the ionised gas and the stars for all the fast-rotating galaxies listed in Table \ref{datatable1}. The dashed line overplotted is a normalized Gaussian distribution with its center at zero and a standard deviation of 15$^{\circ}$, showing the expected scatter.}
\label{ion_star}
\end{figure}

Figure \ref{ion_star} shows that 36$\pm$5\% of fast-rotating galaxies (40 out of 111) have their ionised gas kinematically misaligned from the stars ($\Psi_{\rm ion-star} >$ 30$^{\circ}$), consistent with the percentage of misaligned galaxies found in molecular gas (but with a smaller uncertainty). 
Where both the ionised gas and molecular gas are detected, the measured kinematic misalignments from the stars agree well with each other, with a 1$\sigma$ scatter around the one-to-one relation of 10$^{\circ}$. 
A MW-U test finds no statistically significant evidence that the molecular gas and ionised gas
misalignments from the stars (Figures \ref{mol_star} and \ref{ion_star}) are not drawn from the same underlying parent distribution.

\subsection{Comparison to H{\small I}}
\label{hi}

As part of the \atlas\ survey we collected \hi\ data for all 166 galaxies above declination +10$^{\circ}$ and further than 15 arcmin from Virgo A. Most of these galaxies were observed with the Westerbork Synthesis Radio Telescope (WSRT). Observations are described in detail in \cite{Morganti:2006p1934,Oosterloo:2010p3376} and Serra et al., in prep {\citep[see][for a summary]{Serraconfprop}}. For about 20 galaxies inside the Virgo cluster we use Arecibo data taken as part of the Alfalfa survey \citep{diSeregoAlighieri:2007p3399} instead of WSRT data. We use Alfalfa data to derive an upper limit on the HI mass which is comparable to what we obtain with our WSRT observations.

This \atlas\ \hi\ sample 
 has 29 \hi\ detections of fast-rotators that overlap with the molecular and ionised gas sample with measurable misalignments we consider in this work (13 that overlap with the interferometric CO sample, and a further 16 that overlap with the ionised gas sample).
Despite the large difference in angular resolution between the \hi\ data and the other tracers, in most cases it is possible to estimate a rough kinematic PA of the atomic gas in the region where it overlaps with the CO and ionised gas. 
We find that the atomic and molecular material always appear to be kinematically aligned with one another, suggesting that they share a common origin. This is in agreement with the results of \cite{Morganti:2006p1934}, who find that the \hi\ and ionised gas are always aligned in the SAURON sample (which is a subset of the \atlas\ sample).

As we have established that the ionised, molecular and atomic gas are all likely to be part of the same structure, it is interesting to use the \hi\ data to investigate the connection with \hi\ seen at large radii.
Of the \hi-detected systems that have (molecular or ionised) gas kinematically misaligned from the stars, 86\% (12/14) have extended \hi\ structures, clearly suggesting a source for the misaligned material. These systems are discussed in detail in  \cite{Li:1994p3359} (NGC\,7465), \cite{Verheijen:2001p3449} (NGC\,4026), \cite{Morganti:2006p1934} (NGC\,1023, NGC\, 2685, NGC\,2768, NGC\,7332), \cite{Duc:2007p3378} (NGC\,4694), \cite{Oosterloo:2010p3376} (NGC\,3032, NGC\,4262), and Serra et al., in prep (NGC\,0680, NGC\,3626, UGC\,09519).

Furthermore, 38\% (6/16) of the \hi-detected galaxies that have gas kinematically aligned with the stars also have kinematically disturbed \hi\ at larger radii, with clear suggestions of ongoing accretion from external sources. These are discussed in  \cite{Morganti:2006p1934} (NGC\,4150), \cite{Oosterloo:2010p3376} (NGC\,3384, NGC\,3489) and Serra et al., in prep (NGC\,2778, NGC\,3457, PGC\,028887). 
 
\subsection{Environmental effects}
\subsubsection{Cluster environment}
\label{cluster}
The \atlas\ sample includes a range of galaxy densities, from the Virgo cluster to sparse field environments. Galaxies in clusters are often \hi\ deficient \citep[e.g.][]{diSeregoAlighieri:2007p3399}, and it is thought that the atomic material is removed by ram-pressure stripping and interactions as the galaxies fall into the cluster potential and plow through the hot IC\,M \citep[e.g.][]{Gunn:1972p3253,Giovanelli:1983p3254}. \cite{Kenney:1986p2835}  have shown that spiral galaxies in Virgo are not CO deficient, even when they are \hi\ deficient. This is likely to be because molecular gas is denser and resides deeper in a galaxy potential well, where it will not be as affected by infall into a cluster. 

To first order, the merger rate is enhanced as galaxies fall into clusters, but once the galaxies are virialized the chance of a merger drops off considerably \citep[the large velocity dispersion within the cluster makes mergers reasonably rare; e.g.][]{vanDokkum:1999p3277}. Cluster (and group) centre galaxies are the only exceptions, which are likely to have many merging events, and are typically slow rotators (see Paper VII). 
Paper IV shows that \atlas\ Virgo fast-rotating ETGs are likely to be virialized within the cluster (the line-of-sight velocities are centrally peaked and consistent with being relaxed), and hence their last merger should have occurred several crossing times ago (a few gigayears). 
 
Stripped or free-floating gas is likely to be destroyed and shock heated to the cluster virial temperature as it is captured by the cluster, joining the cluster's hot X-ray halo.  The external channels by which cold gas can be acquired are therefore likely to be shut-off once a galaxy is embedded in a cluster potential, whereas internal stellar mass loss is not impeded. 
This is consistent with the lack of \hi\ detections in Virgo (e.g. \citealt{Giovanardi:1983p3439,diSeregoAlighieri:2007p3399,Oosterloo:2010p3376} and Serra et al., in prep) and with the steep decrease in the frequency of spirals, and increase in the frequency of fast-rotating ETGs in the Virgo cluster core \citep[see][hereafter Paper VII]{Cap2011}

Despite this, as for spirals, the ETG CO detection rate inside the Virgo cluster is much higher than the \hi\ detection rate, and is almost identical to that found in the field (Paper IV). The molecular gas mass fractions (M$_{\rm H_2}$/L$_{K}$) of cluster and field galaxies also appear to be similar.
Paper IV does however show some evidence that the H$_2$-richest ETGs are all low mass, and in poor environments. It is hence interesting to investigate these high- and low-density galaxy populations separately, to see if the origin of the gas is affected by the environment.

When our sample of galaxies is divided into Virgo (defined in Paper I as all the galaxies within a sphere of radius 3.5 Mpc centered on the cluster centre\footnote{Cluster membership is listed in Paper I, Table 5, Column 6, and available online at http://www.purl.org/atlas3d}) and field (everything else), a clear dichotomy in the kinematic misalignment distribution is apparent.

\begin{figure*}
\begin{center}
\includegraphics[scale=0.8,angle=0,clip,trim=0.6cm 0cm 0.7cm 0.7cm]{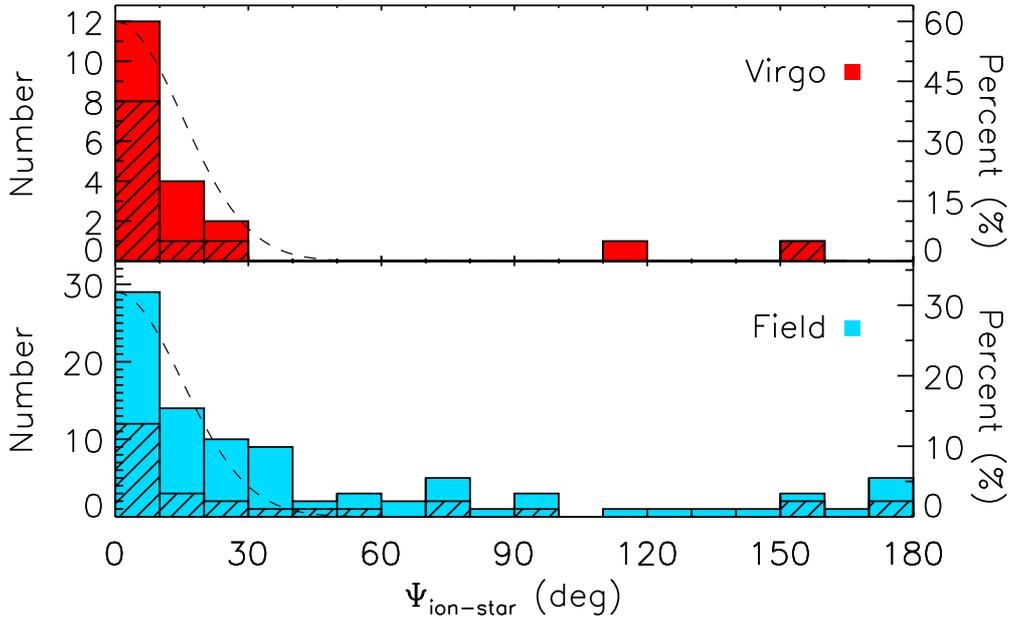}

\end{center}
\caption{\small \textit{Top:} Histogram showing the kinematic misalignment angle between the ionised gas and the stars for all fast-rotator galaxies in Virgo (defined as lying within a sphere of 3.5 Mpc centered on the cluster centre). The hatched area indicates the number of galaxies in each bin that were also mapped in molecular gas. The two outlying galaxies with $\Psi >$30 are NGC\,4262 and NGC\,4694. \textit{Bottom:} As above, but for all fast-rotator galaxies which are not in Virgo, including galaxies in the field and small groups.}
\label{envionstack}
\end{figure*}

In Figure \ref{envionstack}, 42$\pm$5\% (38 out of 91) of the fast-rotating field ETGs have kinematically misaligned ionised gas. However, in the Virgo cluster only 10$\pm$6\% (2 out of 20) of the fast-rotating galaxies have ionised gas kinematically misaligned for the stars ($\Psi_{\rm ion-star} <$ 30$^{\circ}$). If this distribution of alignments in the cluster was the same as in the field one would expect $\approx$8 to 10 galaxies to show misalignments. 

A MW-U test gives a probability that the Virgo and field galaxies are randomly drawn from the same parent galaxy population as 0.3\% ($\approx$3$\sigma$ significance). The same result is found when one considers the kinematic misalignment between the molecular gas and the stars, but with lower number statistics (shown as the hatched area in Figure \ref{envionstack}).

The two kinematically misaligned Virgo galaxies (NGC\,4262 and NGC\,4694) both lie on the outskirts of the cluster and are undergoing mergers/interactions as they fall into the cluster. NGC\,4262 may have recently undergone a close encounter with a nearby galaxy \citep[see][]{Vollmer:2005p3247}. It has a ring of \hi\ surrounding the galaxy 
\citep{Krumm:1985p3248}. 
This system is strongly barred, and the \hi\ shows elliptical orbits which may be the cause of the observed misalignments. 
This ring may be the source of the ionised gas as the kinematic position angles appear to be similar. No CO emission was detected from NGC\,4262 (Paper IV).  NGC\,4694 is undergoing a gas-rich major merger in the cluster outskirts. The CO and \hi\ in this galaxy are in non-equilibrium structures 
offset from the stellar body, towards a dwarf galaxy (\citealt{Duc:2007p3378}; Alatalo et al, in prep.).

\subsubsection{Group environments}

There is observational evidence that preprocessing (through mergers and tidal disruption) in group environments may be important in the formation of ETGs \citep{Zabludoff:1998p3361,Kodama:2001p3362,Helsdon:2003p3363,Kautsch:2008p3360}. Paper VII also shows that local environment density seems to be the most important factor in determining the fraction of spiral galaxies and fast-rotating ETGs, outside of the Virgo cluster core. However, simulations show that galaxies can enter clusters directly from the field \citep[e.g.][]{Berrier:2009p3364}, avoiding any preprocessing in groups, and that therefore any transformation mechanism should also be able to occur in the cluster environment. 

\begin{figure*}

\includegraphics[scale=0.7,angle=0,clip,trim=0cm 0cm 0cm 0cm]{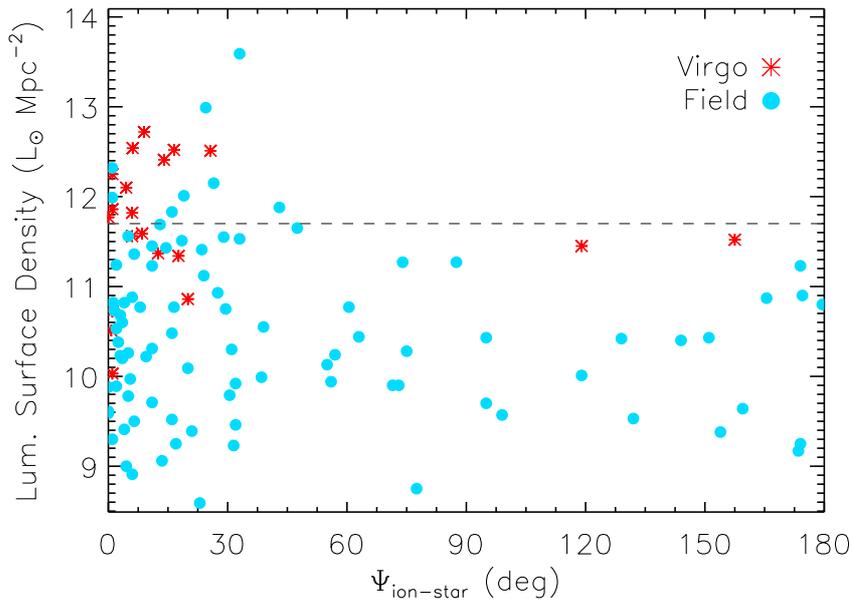}

\caption{\small The kinematic misalignment angle between the ionised gas and the stars for fast-rotator galaxies, plotted against the local luminosity surface density within cylinders of depth 600 \kms\ in the redshift direction, with an angular size adapted to include the 3rd nearest neighbor. Virgo galaxies are plotted with red stars, and field/group galaxies with solid blue circles. The dashed line is a guide to the eye, at the suggested critical density of 10$^{11.7}$ L$_{\odot}$ Mpc$^{-2}$. The error on each kinematic misalignment angle measurement is $\approx$15$^{\circ}$.}
\label{localenv_ion}
\end{figure*}

It is possible with the data we possess to examine if it is the local density or the global environment that drive the observed kinematic misalignment correlations. To do this, we utilize a luminosity surface density estimator ($I_3$) based on a redshift cylinder with a depth of 600 \kms, and an angular size adapted to include the 3rd nearest neighbor, as presented in \cite{Cap2011} (Paper VII; Table 2, Column 7 and available online at http://www.purl.org/atlas3d). This estimator should pick out locally dense environments, such as groups, as well as substructures within the Virgo cluster. We plot this density indicator against the kinematic misalignment of the ionised gas and stars in Figure \ref{localenv_ion}. 

In the densest environments, above a critical local luminosity surface density of $\approx$ 10$^{11.7}$ L$_{\odot}$ Mpc$^{-2}$, all fast-rotators in both the field and cluster have ionised gas kinematically aligned with the stars, while a wide range of misalignments exist at lower densities. The two misaligned systems in the Virgo cluster are below this luminosity surface density (they are thought to be recently accreted systems, still undergoing mergers in the cluster outskirts; see Section \ref{cluster}). This result does not change if one plots number surface density ($\Sigma_3$; see Paper VII, Table 2, Column 4) rather than luminosity surface density, with the critical number density being $\approx$ 20 Mpc$^{-2}$. A MW-U test gives a (a posteriori) probability of $\approx$1\% that the kinematically aligned galaxies above (and all the galaxies below) this critical density are randomly drawn from the same 
parent population. If one considers the larger scale $I_{10}$ and $\Sigma_{10}$ environment indicators from Paper VII the cluster/field result discussed in Section \ref{cluster} is recovered.

The field galaxies that exist above this critical density are NGC\,0524, NGC\,0680, NGC\,3379,  NGC\,3412, NGC\,5353, NGC\,5355, NGC\,5379 and PGC\,042549, all of which are in known galaxy groups. All but one of these systems (which were observed in our \hi\ survey) are \hi\ poor (NGC\,0680 is the exception, it is a major merger remnant, and has kinematically disturbed \hi\ extending between several of its group members; see Paper IX and Serra et al., in prep). At least two of these groups have extended X-ray halos (NGC\,524, \citealt{Romer:2000p3365}; NGC\,5353/NGC\,5355, \citealt{OSullivan:2001p3380}).

\subsection{Mass dependence}
\label{masseffect}
\begin{figure*}
\includegraphics[scale=0.7,angle=0,clip,trim=0cm 0cm 0cm 0cm]{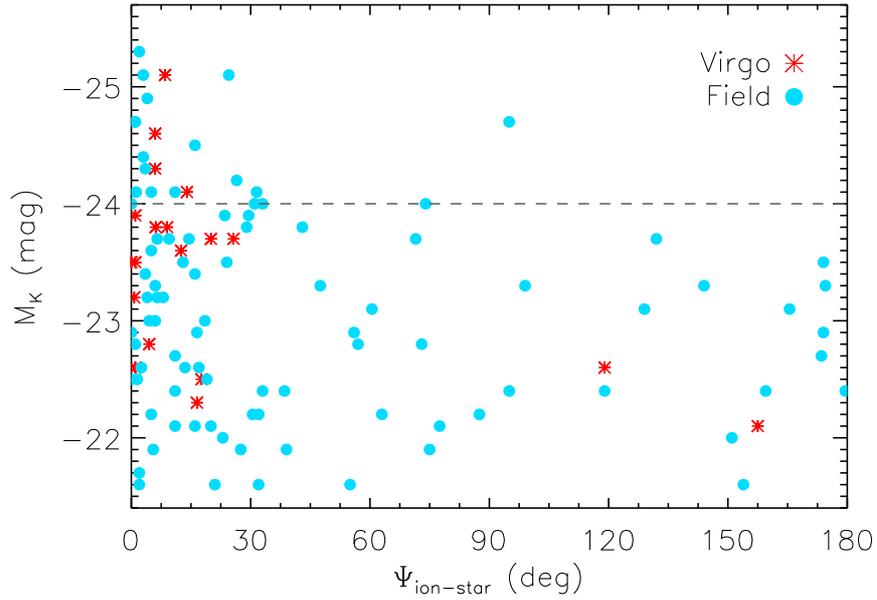}

\caption{\small The kinematic misalignment angle between the ionised gas and the stars for fast-rotator galaxies, plotted against the total absolute $K_{\rm s}$-band magnitude of the galaxy from 2MASS. Distances to the galaxies are taken from Paper I. Virgo galaxies are plotted with red stars, and field/group galaxies with solid blue circles. The dashed line is a guide to the eye, at the suggested critical magnitude of M$_K$=$-$24 mag. The error on each kinematic misalignment angle measurement is $\approx$15$^{\circ}$.}
\label{mk_ion}
\end{figure*}

Whenever one sees an effect which appears correlated with environment, it is important to check if the observed relation is in fact driven by galaxy mass. 
The mass of a galaxy can affect the gas content in various ways. A deeper potential can allow a galaxy to retain gas that would be lost due to supernovae and galactic winds in a lower mass system. Massive ETGs can also have X-ray halos, which can prevent external gas (especially in cold flows) from entering the galaxy.  Figure \ref{mk_ion} shows the dependence of the ionised gas-star kinematic misalignment of fast-rotators on absolute \mbox{$K_{\rm s}$-band} magnitude, an observable proxy for stellar mass. Distances to the galaxies are taken from Paper I. One can see a clear lack of misaligned fast-rotating galaxies at M$_{K}\ltsimeq\,-24$. The only kinematically misaligned galaxy brighter than this threshold is NGC\,2768, which is a large elliptical galaxy in a relatively isolated environment, that appears to be accreting from a nearby cloud of \hi\ \citep{Morganti:2006p1934,Crocker:2008p946}. A MW-U test gives a $\approx$2\% chance that the galaxies with M$_K \ltsimeq -24$ are drawn from the same distribution of misalignment angles as the fainter galaxies.

\begin{figure*}
\includegraphics[scale=0.7,angle=0,clip,trim=0cm 0cm 0cm 0cm]{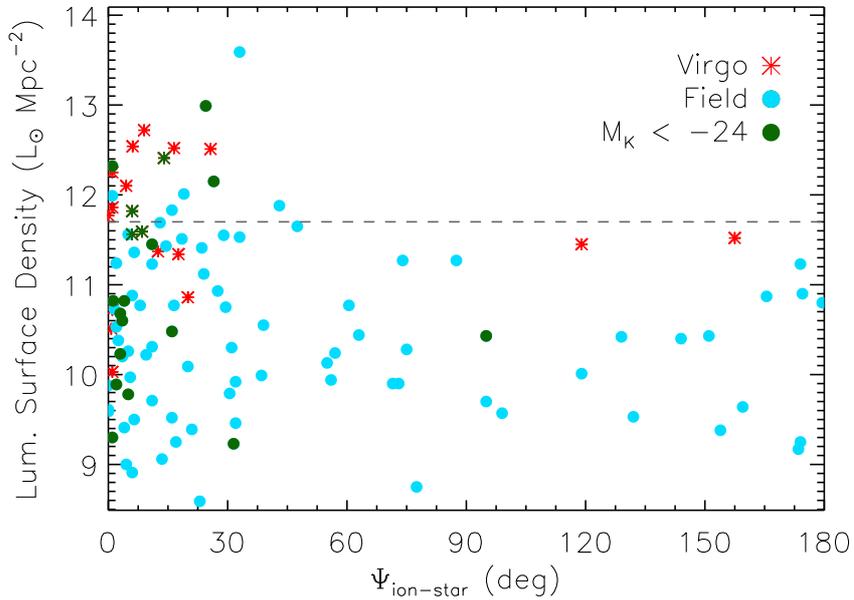}
\caption{\small As Figure \protect \ref{localenv_ion}, but galaxies with M$_{\rm K} <$  $-$24 mag are plotted with dark green stars (for Virgo galaxies) and solid dark green circles (for field/group galaxies). The error on each kinematic misalignment angle measurement is $\approx$15$^{\circ}$.}
\label{dens_ion_wxray}
\end{figure*}

We re-plot Figure \ref{localenv_ion} with the galaxies with M$_K < -$24 mag highlighted in green (Figure \ref{dens_ion_wxray}). Interestingly, the fast-rotators above the proposed critical luminosity are not all the same systems identified previously as being in dense environments, in fact, most are different. 
This is likely to be because the brightest and most massive galaxies in dense environments are usually slow rotators (Paper VII). 
It seems that for fast-rotating galaxies both mass \textit{and} environment can be important independent parameters in determining the origin of the molecular and ionised gas.

\begin{table*}
\caption{Synopsis of the statistics presented in this paper}
\begin{tabular*}{1\textwidth}{@{\extracolsep{\fill}}r r r r r r r r}
\hline
Category & Number & Mol. Detection & Ion Detection & \# Mapped Mol &  \# Mapped Ion &  Misaligned Mol &  Misaligned Ion \\
(1) & (2) & (3) & (4) & (5) & (6) & (7) & (8)\\
\hline
Fast-Rotators & 224 & 24$\pm$3\% & 67$\pm$\hspace{4pt}3\% & 36 & 111 & 23$\pm$\hspace{4pt}7\% & 36$\pm$\hspace{4pt}5\% \\
Slow-Rotators & 36 & 6$\pm$4\% & 86$\pm$\hspace{4pt}6\% & 2 & 22 & 50$\pm$35\% & 68$\pm$10\%\\
Field FR & 202 & 22$\pm$3\% & 73$\pm$\hspace{4pt}3\% & 26 & 91 & 28$\pm$\hspace{4pt}9\% & 42$\pm$\hspace{4pt}5\%  \\
Virgo FR & 58 & 21$\pm$5\% & 47$\pm$\hspace{4pt}6\% & 10 & 20 &  10$\pm$\hspace{4pt}9\% & 10$\pm$\hspace{4pt}6\%\\
$I_{3}$ $<$ 10$^{11.7}$ FR & 209 & 22$\pm$3\% & 67$\pm$\hspace{4pt}3\% & 35 & 93 & 24$\pm$\hspace{4pt}7\% & 44$\pm$\hspace{4pt}5\%\\
$I_{3}$ $\geq$ 10$^{11.7}$ FR & 51&  16$\pm$5\% & 69$\pm$10\% & 1 & 18 & 0\% & 16$\pm$\hspace{4pt}9\%\\
M$_{\rm Ks}$ $\geq$ -24 FR & 191 & 24$\pm$3\% & 65$\pm$\hspace{4pt}3\% & 29 & 92 & 39$\pm$\hspace{4pt}9\% & 44$\pm$\hspace{4pt}7\% \\
M$_{\rm Ks}$ $<$ -24 FR & 33 & 27$\pm$8\% & 80$\pm$\hspace{4pt}7\% & 7 & 19 & 14$\pm$13\% & 5$\pm$\hspace{4pt}5\%\\
\hline
\end{tabular*}
\parbox[t]{1 \textwidth}{ \textit{Notes:} Column 1 lists the categories of galaxies we discuss in this paper. FR denotes fast-rotators only. $I_3$ is our local the luminosity surface density estimator, based on a redshift cylinder with a depth of 600 \kms, and an angular size adapted to include the 3rd nearest neighbor, as presented in \cite{Cap2011} (Paper VII; Table 2, Column 7 and available online at http://www.purl.org/atlas3d). M$_{Ks}$ is the absolute $Ks$-band total magnitude. Column 2 lists the number of galaxies from the \atlas\ survey that fall in this class. Columns 3 and 4 list the molecular gas detection rates (taken from \citealt{Young2010}) and ionised gas detection rates (above an integrated equivalent width of 0.02 \AA\ in either H$\beta$ or [OIII]) for these subsamples. Columns 5 and 6 state the number of galaxies in this category mapped in molecular gas, and those with measurable kinematic PA's in their ionised gas maps. Columns 7 and 8 state the percentage of mapped galaxies which have misaligned molecular, and ionised gas respectively ($\Psi$ $>$ $30^{\circ}$).  }

\label{statstable}

\end{table*}

 \section{Discussion}
 \label{discuss}
 
The results presented in the previous section shed light on the origin of the molecular and ionised gas in local ETGs, but they pose additional puzzles that require further discussion.

\subsection{The link between the gaseous phases}
\label{ion_discuss}
Our results suggest that the molecular, atomic and ionised gas form a single structure, and have undergone a common evolution \citep[as previously found for atomic and ionised gas by][]{Morganti:2006p1934,Oosterloo:2010p3376} at least in fast-rotating galaxies. 
It would be difficult to maintain two kinematically distinct phases of the ISM for any length of time, given the collisional nature of gas (unless they were widely separated in radius, e.g. separate rings).  
For instance, let us consider a galaxy which has stellar mass loss ongoing, and has built an aligned gas reservoir. 
If this galaxy acquires misaligned gas from an external source, the two systems may interact.
Collisions between two components with different angular momentum vectors should cause some of the gas to lose its angular momentum and fall to the centre resulting in a more compact gas distribution. A search for such a signature will be presented in a later work in this series.

The ionised gas in these galaxies will also be coupled to the properties of the broader ISM (due to its dissipative nature).
The dominant ionisation mechanism in ETGs is thought to be irradiation by old stellar populations \citep[e.g. post-asymptotic giant branch stars;][]{Binette:1994p3381,Sarzi:2009p2933}, which are uniformly distributed throughout the galaxy, hence ionised gas should be produced at the interface of any part of the cold ISM, and the ionised gas would have the same angular momentum as the material it is produced from. 

{The difference in angular resolution (and field of view) between our CO and \hi\ observations (and to a lesser extent our ionised gas observations) could have some effect on these conclusions. For instance warps are often seen in outer \hi\ disks around galaxies of all types \citep[e.g.][]{GarciaRuiz:2002p3468}. However the agreement between the CO (and ionised gas) in the inner parts, and the \hi\ at larger radii would be hard to manufacture by chance. Although challenging, higher resolution \hi\ observations of the inner parts of these galaxies could be used to remove this uncertainty, and further establish the link between these phases. }

\subsection{The importance of external accretion}

\subsubsection{Slow-rotators}
In slow rotators there is clear evidence that their ionised gas is primarily acquired externally. The kinematic misalignments are statistically consistent with a flat distribution, with no significant increase in the number of galaxies with exactly co- or counter-rotating gas. This suggests that accretion and mergers dominate the gas supply in slow-rotating galaxies. If the source of ionization in these galaxies is long lasting (e.g. irradiation by old stellar populations) then this gas could even be left-over from the last merger that created the slow-rotator. 

Slow rotators are thought to be formed from progenitors that have undergone multiple major mergers (Paper III). Simulations suggest an average of 3 major mergers per slow-rotating galaxy, and multiple minor mergers \citep[][Paper VIII]{Khochfar2011}. Slow rotators are predominately massive (Paper III), and sit at the centre of their group and cluster environments (Paper VII), which is likely to enhance their merger rate \citep[e.g.][]{vanDokkum:1999p3277}.
Our results provide additional evidence that mergers and/or accretion are important in the evolution of slow rotators.

A few questions do remain. For instance there is no statistically significant evidence for any of these galaxies having generated aligned ionised gas from stellar mass loss. The stars in these systems are predominantly old (e.g. \citealt{Kuntschner:2010p3471}; McDermid et al., in prep), and hence the mass loss rate from the stellar population will be low. Any mass that is lost from these stars into the ISM may be kept hot by the large velocity dispersion of these pressure supported systems, and may join the X-ray halo.

Furthermore, the lack of a peak in the misalignment distribution for co- and counter-rotating gas suggests the timescale required for the ionised gas in these systems to relax into the orbital plane is long (especially as the ionised gas is likely to be long lived). This may be because these systems are in general very round (Paper III), and hence the gravitational torques will be lower than in disky systems. Slow rotators can also be triaxial (Paper II \& Paper III), allowing gas in polar configurations to be stable. 

\subsubsection{Fast rotators}

The molecular and ionised gas in $\approx$36\% of the galaxies in our sample are kinematically misaligned from the stellar body, and hence are likely to have an external origin. There are ways in which the angular momentum of gas in the central regions of galaxies can be altered \citep[e.g. bars which produce resonant orbits and branching; see][]{Wada:1995p3443}. However the kinematic alignment with the atomic gas at larger scales suggests this is not a dominant effect. 
In addition, 36\% is likely a lower limit to the importance of externally acquired material. Assuming that gas from mergers and accretion enters galaxies isotropically then one would expect 
on average $\approx$17\% of galaxies to exhibit aligned kinematics (within 30$^{\circ}$) by chance (30$^{\circ}$/180$^{\circ} \approx$  0.17), despite the gas having an external origin. 

The assumption that gas from mergers and accretion enters galaxies isotropically is a conservative one. In fact one can imagine scenarios where the gas accretion is not isotropic, but in fact preferential results in aligned gas. One such example would be a major merger that imparts a large amount of angular momentum to the stars in the remnant, forcing them to rotate in the plane of the merger. Gas that is thrown off in the merger may fall back into the remnant over many gigayears, and would likely fall in preferentially aligned. Some simulations have predicted that minor mergers may preferentially occur with the angular momentum vector of the accreted system aligned along the major axes of the dark matter halo \citep[e.g.][]{Deason:2011p3441}. If such processes are ongoing, then we will underestimate the importance of external accretion in this work. How isotropic the accretion of gas from major mergers and cold-mode accretion onto ETGs truly is could be tested further in cosmological hydrodynamic simulations. 

If a large amount of external gas is present within 
a galaxy, the gas depletion timescale through star-formation will become larger than the relaxation time (the time it takes for gas structures to settle into the equatorial plane). In this case one would expect to find 50\% of the galaxies to have co-rotating gas, and 50\% to show counter-rotation. The lack of a large peak at a kinematic misalignment of 180$^{\circ}$ however suggests that this criterion is seldom fulfilled, or that relaxation into the equatorial plane is not a symmetric process.

\cite{Clemens:2010p3473} compared the dust production rate in the envelopes of evolved AGB stars with a constraint on the total dust mass in Virgo cluster ETGs (from Herschel observations). They find that passively evolving ETGs (defined by having Spitzer mid-infrared spectra consistent with passively evolving stellar populations) are not detected by Herschel, and thus the dust destruction timescale in passive ETGs must be short. They conclude that passive ETG's with dust are thus likely to have acquired it from an external source. As dust is thought to be important to shield nascent molecular clouds, and catalyze the conversion of \hi\ into H$_{2}$ molecules \citep[e.g.][]{Duley:1993p3469}, perhaps a lack of dust could prevent some of our sample galaxies from regenerating their molecular reservoirs from internal sources. 

External accretion/mergers appear to only be important in the field, where at least 42\% of galaxies have a significant kinematic misalignment between the molecular or ionised gas and the stars. The molecular and ionised gas are nearly always consistent with a purely internal origin in three (partially overlapping) populations:  fast-rotating Virgo cluster galaxies, the densest group environments we sample, and in the highest luminosity fast-rotators.

As shown in Section \ref{hi}, there is clear evidence that at least 6 of the field ETGs which are kinematically aligned have gas which is in fact externally accreted. Including this evidence increases the percentage of field galaxies which have externally accreted gas to at least 46\%. 
\cite{Knapp:1985p3440} have shown that the \hi\ masses of early-type galaxies are unrelated to the galaxy luminosity, unlike in spirals. This led them to predict that external sources provide \textit{all} the \hi\ in field ETGs. 
\cite{Morganti:2006p1934} and \cite{Oosterloo:2010p3376} have also shown that in almost all cases where \hi\ is present in ETGs, there are some signs that it has been externally accreted. 
As it appears that the gaseous phases are closely linked, this suggests that the molecular and ionised gas in \hi-detected ETGs frequently has an external origin, even if it is kinematically aligned. Hence the true importance of external accretion is likely underestimated in field environments.

If this is so, then we are forced to conclude that mergers and accretion must not proceed isotropically, or relaxation of misaligned gas happens over a short timescale, and preferentially in a prograde direction. It is also conceivable that retrograde gas survives for a shorter length of time (due to interactions with the hot halo of the galaxy, or similar).
Due to the different angular resolutions (and hence spatial scales) probed by the \hi\ and CO data however, these conclusions require further investigation. {To first order, one would expect the evidence for external \hi\ accretion to become stronger if one were able to investigate the kinematics and morphology of the \hi\ at higher spatial resolution.} A more in depth treatment of these issues will be included in a future work in this series.  

\subsection{The effect of group and cluster environments}
The striking result that galaxies in the Virgo cluster (and other dense environments) almost always have kinematically aligned gas suggests that dense environments affect the accretion and/or generation of ionised and molecular material. 
The challenge is thus to find a consistent theory that can explain the environmental dependence of the kinematic misalignments, while simultaneously accounting for the identical detection rates (and mass fractions) of the molecular gas in the field and clusters (Paper IV).

Paper IV has shown that the CO-detected galaxies in Virgo are virialised, and hence have been in the cluster for several gigayears. 
The average gas depletion timescale for field galaxies in our CO-detected sample is $\approx$1-2 Gyr, assuming that the gas forms stars at the rate predicted by either the Schmidt-Kennicutt relation of \cite{Kennicutt:1998p3214} or a constant star formation efficiency as in \cite{Bigiel:2008p1868} (See Alatalo et al., in prep). 
Recent studies have found little evidence that these ETGs deviate from these star-formation relations, and hence morphological quenching may not play a large role in the prolonging the lifetime of the the gas in majority of these systems (see \citealt{Shapiro:2010p2932}, \citealt{Crocker:2011p3291} and Alatalo et al, in prep). 

If the ETGs present in clusters today were like current field galaxies when they fell in, then they should have exhausted the vast majority of their molecular gas by now (due to star formation) and should not be detectable anymore. This implies that the ETGs detected in Virgo today either must have been systematically more molecular gas-rich than is typical now when they fell in, or they must have regenerated their gas from an internal source recently. Of course, if ETGs were more molecular gas-rich in the past, then either all the gas must have been kinematically aligned (contrary to today's field ETGs) or it must have somehow re-aligned upon entering the cluster. Both possibilities however seem rather contrived, so the internal replenishment hypothesis seems more plausible. \citep{Skillman:1996p3470} showed that spiral galaxies in the inner parts of the Virgo cluster have higher gas-phase metallicities than those in the field. They suggested that these systems have been evolving as `closed-boxes', undiluted by metal poor gas infall, providing some support for this hypothesis.

It is of course possible that the molecular (and ionised) gas in ETGs in Virgo all comes from cooled stellar mass loss. This would however require that ETGs which fall into the cluster exhaust their pre-existing gas rapidly, so that any kinematically misaligned gas is destroyed, and then that the right percentage of the ETG population regenerate a molecular gas reservoir.

Figure \ref{localenv_ion} suggests that high local density environments also result in aligned gas.  We speculate that preprocessing in groups, before galaxies enter clusters, could perhaps be important in explaining the kinematic alignment of the gas. 
As galaxies fall into groups, or transit through the outskirts of clusters, merger rates are enhanced \citep[e.g.][]{vanDokkum:1999p3277}. Starbursts triggered by these violent interactions can quickly deplete the cold molecular gas, which would rapidly remove any evidence of misaligned material. 
The galaxies in \hi-poor groups and clusters would then be unable to re-accrete external gas as they settle into the group/cluster potential and become virialized. Stellar mass loss can then regenerate an aligned gas reservoir over time. 

The principal difficulty of this explanation is that it requires the gas generation efficiency to be higher in clusters than in field galaxies. Cluster galaxies have the same molecular gas detection rate and mass fractions as field galaxies, but all this molecular gas must be generated internally. 
About 11\% of galaxies in the field have a substantial molecular gas reservoir kinematically aligned with the stellar component (i.e. half of the detections), while $\approx$20\% of galaxies in Virgo have kinematically aligned gas (i.e. almost all of the detections). 
The \atlas\ ionised gas detection rate is $\approx$73\% in the field, but only $\approx$48\% in the cluster. If this ionised gas is cooling to become molecular, then it must do so with a much greater efficiency in the cluster environment.

Some studies of the formation of molecular clouds \citep[e.g.][]{Heitsch:2008p3258} have suggested that they are most likely to form at the intersection of gas flows or shocks, such as those found in the spiral arms and bars of disc galaxies. Conceivably, if such conditions are found more often in cluster galaxies, this could result in a higher molecular cloud formation efficiency. 
The morphological classification of the \atlas\ galaxies presented in Paper II reports a slightly enhanced barred fraction in the Virgo cluster (36$\pm$4\% versus 27$\pm$2\% in the field). 
Almost all of the CO-detected galaxies in Virgo and dense groups that are aligned have their molecular gas distributed in rings, spirals, or within bars (see Alatalo et al, in prep). These structures associated with galaxy resonances can perhaps provide an environment with suitable gas flows or shocks, that could concentrate any diffuse gas present (from stellar mass loss) and allow it to cool and form molecular clouds. The small difference in barred fraction within the cluster is unlikely to be enough to explain the  apparent enhanced molecular gas formation efficiency however.

Paper VII revisited the morphology-density relation \citep{Dressler:1980p2456} for the galaxies in our sample using the kinematic classification presented in Paper III. It showed that the fraction of spirals decreases, and the fraction of fast-rotating ETGs increases as you approach a group or cluster core.
This can be interpreted as circumstantial evidence that spiral galaxies can be transformed into fast-rotating ETGs as they settle into a group or cluster.  Therefore, it is possible that some of the ETGs we detect in the group and cluster environments actually fell in as spirals, which were then subsequently transformed into early-types. We would then be detecting the remnant of their star-forming gas, that would naturally be co-rotating (counter-rotating molecular gas is very rare in spirals). 
This explanation would however require accepting that the molecular gas mass fraction is coincidentally similar inside and outside the cluster. 
The CO-detected ETGs in Virgo are also statistically indistinguishable from the other ETGs in Virgo in their brightness at $K$-band, specific angular momentum ($\lambda_{\rm R}$, from Paper III) , dynamical mass or stellar velocity dispersion (both from Cappellari et al, in prep), alpha-element enhancement and stellar metallicities (at least in single stellar population analyses, McDermid et al., in prep). 
Nevertheless, we feel that this possibility is worthy of further study, as it may provide direct observational evidence of galaxies undergoing morphological change due to cluster environments. 

All of the mechanisms discussed above fail to satisfactorily explain the identical detection rate and molecular gas fractions found inside and outside of Virgo.  
The low number of detected ETGs means that the detection rates and molecular gas fractions in the cluster may be identical by chance, and of course other clusters may be very different. Clearly observations of other clusters are required to resolve this ambiguity, and pinpoint which, if any, of the above mechanisms are dominant.

\subsection{The effect of galaxy mass}

Figure \ref{mk_ion} shows that the kinematic misalignments of the molecular and ionised gas in fast-rotating ETGs also seem to be effected by the mass of the host galaxy. Furthermore, this effect appears to be independent of the environmental effect discussed above (see Figure \ref{dens_ion_wxray}). The processes resulting in these kinematic alignments 
must thus act on single galaxy scales.

Massive galaxies are better able to contain material that would normally be lost during energetic events (such as supernovae/galactic winds), perhaps resulting in a larger pool of aligned material that can later cool.
This process, however, can only create aligned gas, and thus leaves us unable to explain the lack of externally accreted gas in these massive fast-rotating galaxies.

AGN feedback is one of the possible mechanisms which could explain this lack of kinematically misaligned gas. \cite{Springel:2005p3425} have shown in simulations that galaxy mergers and accretion events that cause nuclear gas inflow can trigger AGN feedback, quenching star formation and gas accretion on a short timescale, particularly in massive galaxies. 
AGN feedback triggered by merging events could disrupt and destroy misaligned gas reservoirs as they form in the most massive galaxies. Gas from stellar mass loss can avoid such a fate, as it will not flow into the centre in great quantities due to angular momentum conservation, and hence will not fuel an AGN.

Massive ETGs are also known to host hot, X-ray emitting gas halos \citep{Forman:1985p3426}. These hot halos can heat infalling cold gas clouds, preventing the gas from reaching the galaxy centers.  In a similar way, as mentioned above, virial shocks can create a halo mass threshold, above which incoming gas will always shock heat to the virial temperature, joining the hot halo \citep{Birnboim:2003p3427}. 
This material is virialised in the hot halo, and likely forced into an aligned configuration by interaction with the rest of the halo gas. It may then cool in an identical way to stellar mass loss, producing a kinematically aligned gas reservoir. 

Both of the halo effects described above are reported to occur primarily in galaxies of mass $\gtsimeq10^{11}$ M$_{\odot}$, a good match to our apparent threshold (M$_K$ = $-24$ mag, or 8$\times$10$^{10}$ $L_{\odot,K}$ ). Once again, however, the fact that our molecular gas detection rate is independent of mass makes explaining these results difficult. Massive galaxies may well retain a large reservoir of kinematically aligned hot gas (from internal or external sources), but in order to achieve a uniform detection rate as a function of mass the cooling rate from the halos would have to be enhanced by just the right amount to balance the reduction due to the lack of accretion events. It is possible that the star formation efficiency in these systems could be lower, perhaps due to morphological quenching, allowing galaxies that do have gas to be detectable for longer. 

Slow-rotating galaxies in general have a similar mass to this population of aligned fast-rotators ($>$3x10$^{10}$ M$_{\odot}$; Paper III), but have gas that is mainly from external sources. This suggests, as discussed in Papers III and VIII, that slow-rotators have evolved differently to fast-rotators, with a greater number of external events, such as mergers, shaping their properties.

\section{Conclusions}
In this work we have demonstrated that a large proportion of the gas found in local early-type galaxies is likely supplied by external processes. 
Gas kinematically misaligned with respect to the stars is common, indicating an external origin for the gas in $\gtsimeq$36\% of all fast-rotating ETGs. 
We have also shown that the ionised, atomic and molecular gas in local ETGs are linked, always having similar kinematics and thus presumably sharing a common origin. In the field, $\approx$42\% of galaxies have kinematically misaligned ionised gas, confirming that mergers and accretion play a vital role in supplying gas to field ETGs. Slow rotators are not generally detected in molecular gas, but show a flat distribution of ionised gas kinematic misalignments, suggesting they obtain ionised gas primarily through mergers and/or accretion. In many cases, fast-rotating galaxies with kinematically aligned molecular and ionised gas have \hi\ distributions that again suggest an external origin for the gas. This suggests that external gas sources dominate in the field (as has been suggested by previous authors).

In the Virgo cluster, the molecular and ionised material in fast-rotators is nearly always kinematically aligned with the bulk of the stars, pointing to gas supplied by purely internal processes. 
Fast-rotators in dense groups also appear to always have aligned gas kinematics, suggesting that the local environment is important in understanding the environmental dependance of the gas origins. 
Given the results of Paper IV, indicating similar molecular gas detection rates and mass fractions in clusters and the field, this sets constraints on potential mechanisms for generating and influencing molecular and ionised gas in cluster environments. 

Molecular and ionised gas in the most massive fast-rotating galaxies also appear to always be kinematically aligned with the stars, independent of environment, suggesting that the alignment can be caused by galaxy scale processes which reduce the probability that cold, kinematically misaligned gas can be accreted onto the galaxy (e.g. AGN feedback, the ability to host a hot X-ray gas halo, or a halo mass threshold). This also supports the picture (discussed in Papers III and VIII) that fast-rotating galaxies (which are always aligned at high mass) have a different formation mechanism to slow-rotators (who at similarly high masses mainly have externally supplied ionised gas).

We tentatively suggest that preprocessing in groups may help to explain the environmental dichotomy in the origin of the gas. Merger-induced starbursts in groups and cluster outskirts could consume any kinematically misaligned molecular gas that is present, and leftover ionised material would be ram pressure-stripped. Once galaxies settle into a cluster or \hi-poor group, external gas accretion and mergers are suppressed, allowing stellar mass loss to regenerate a kinematically aligned gas reservoir.  It is possible that features such as bars and rings could funnel dust and gas lost by stars to the centre of the galaxy, or collect them together, and could explain the greater efficiency with which galaxies in dense environments must recreate their dense gas reservoirs. Alternatively, we could be detecting the remnant gas left over from the morphological transformation of spiral galaxies into ETGs as they enter the cluster and group environments. Both of these possibilities, however, fail to explain why the detection rate of molecular gas (and the molecular gas mass fractions) are similar inside and outside of the Virgo cluster.
 
More work to pin down the mechanism(s) creating and/or aligning the gas is clearly required to unambiguously determine which, if any, of these effects is dominant.

It would also be beneficial to extend this sort of analysis to other clusters and groups, to see if the results reported hold true in yet denser environments. The Fornax cluster in the southern hemisphere and the Coma cluster in the north are obvious nearby targets, which should become accessible once new facilities such as the Large Millimeter Telescope (LMT) and the Atacama Large Millimeter/sub-millimeter Array (ALMA) come online.

\vspace{10pt}
\noindent \textbf{Acknowledgments}
TAD would like to thank Genevieve Graves and Bill Mathews for useful discussions. 
MC acknowledges support from a Royal Society University Research Fellowship.
This work was supported by the rolling grants `Astrophysics at Oxford' PP/E001114/1 and ST/H002456/1 and visitors grants PPA/V/S/2002/00553, PP/E001564/1 and ST/H504862/1 from the UK Research Councils. RLD acknowledges travel and computer grants from Christ Church, Oxford and support from the Royal Society in the form of a Wolfson Merit Award 502011.K502/jd. RLD also acknowledges the support of the ESO Visitor Programme which funded a 3 month stay in 2010.
SK acknowledges support from the the Royal Society Joint Projects Grant JP0869822.
RMcD is supported by the Gemini Observatory, which is operated by the Association of Universities for Research in Astronomy, Inc., on behalf of the international Gemini partnership of Argentina, Australia, Brazil, Canada, Chile, the United Kingdom, and the United States of America.
TN and MBois acknowledge support from the DFG Cluster of Excellence `Origin and Structure of the Universe'.
MS acknowledges support from a STFC Advanced Fellowship ST/F009186/1.
NS and TD acknowledge support from an STFC studentship.
The authors acknowledge financial support from ESO. 
This paper is partly based on observations carried out with the IRAM Thirty Meter Telescope and IRAM Plateau de Bure Interferometer. IRAM is supported by INSU/CNRS (France), MPG (Germany) and IGN (Spain).
Support for CARMA construction was derived from the states of California, Illinois, and Maryland, the James S. McDonnell Foundation, the Gordon and Betty Moore Foundation, the Kenneth T. and Eileen L. Norris Foundation, the University of Chicago, the Associates of the California Institute of Technology, and the National Science Foundation. Ongoing CARMA development and operations are supported by the National Science Foundation under a cooperative agreement, and by the CARMA partner universities. 
We acknowledge use of
the NASA/IPAC Extragalactic Database (NED) which is operated
by the Jet Propulsion Laboratory, California Institute of Technology,
under contract with the National Aeronautics and Space Administration.

\bsp
\bibliographystyle{mn2e}

\begin{thebibliography}{95}
\expandafter\ifx\csname natexlab\endcsname\relax\def\natexlab#1{#1}\fi

\bibitem[{Bacon {et~al}\mbox{.}(1995)Bacon, Adam, Baranne, Courtes, Dubet,
  Dubois, Emsellem, Ferruit, Georgelin, Monnet, Pecontal, Rousset, \&
  Say}]{Bacon:1995p3377}
Bacon R. {et~al.}, 1995, A\&AS, 113, 347

\bibitem[{Bacon {et~al}\mbox{.}(2001)Bacon, Copin, Monnet, Miller,
  Allington-Smith, Bureau, Carollo, Davies, Emsellem, Kuntschner, Peletier,
  Verolme, \& de~Zeeuw}]{Bacon:2001p1477}
---, 2001, MNRAS, 326, 23

\bibitem[{Baldry {et~al}\mbox{.}(2004)Baldry, Glazebrook, Brinkmann,
  Ivezi{\'c}, Lupton, Nichol, \& Szalay}]{Baldry:2004p3398}
Baldry I.~K., Glazebrook K., Brinkmann J., Ivezi{\'c} {\v Z}., Lupton R.~H.,
  Nichol R.~C., Szalay A.~S., 2004, ApJ, 600, 681

\bibitem[{Barnes(2002)}]{Barnes:2002p2041}
Barnes J.~E., 2002, MNRAS, 333, 481

\bibitem[{Beck, Turner \& Kloosterman(2007)Beck, Turner, \&
  Kloosterman}]{Beck:2007p2941}
Beck S.~C., Turner J.~L., Kloosterman J., 2007, AJ, 134, 1237

\bibitem[{Berrier {et~al}\mbox{.}(2009)Berrier, Stewart, Bullock, Purcell,
  Barton, \& Wechsler}]{Berrier:2009p3364}
Berrier J.~C., Stewart K.~R., Bullock J.~S., Purcell C.~W., Barton E.~J.,
  Wechsler R.~H., 2009, ApJ, 690, 1292

\bibitem[{Bigiel {et~al}\mbox{.}(2008)Bigiel, Leroy, Walter, Brinks, de~Blok,
  Madore, \& Thornley}]{Bigiel:2008p1868}
Bigiel F., Leroy A., Walter F., Brinks E., de~Blok W. J.~G., Madore B.,
  Thornley M.~D., 2008, AJ, 136, 2846

\bibitem[{Binette {et~al}\mbox{.}(1994)Binette, Magris, Stasi{\'n}ska, \&
  Bruzual}]{Binette:1994p3381}
Binette L., Magris C.~G., Stasi{\'n}ska G., Bruzual A.~G., 1994, A\&A, 292, 13

\bibitem[{Birnboim \& Dekel(2003)}]{Birnboim:2003p3427}
Birnboim Y., Dekel A., 2003, MNRAS, 345, 349

\bibitem[{Bock {et~al}\mbox{.}(2006)Bock, Bolatto, Hawkins, Kemball, Lamb,
  Plambeck, Pound, Scott, Woody, \& Wright}]{Bock:2006p2806}
Bock D. C.-J. {et~al.}, 2006, Ground-based and Airborne Telescopes. Proc. SPIE,
  6267, 13

\bibitem[{Bregman \& Parriott(2009)}]{Bregman:2009p2870}
Bregman J.~N., Parriott J.~R., 2009, ApJ, 699, 923

\bibitem[{Cappellari \& Copin(2003)}]{Cappellari:2003p3284}
Cappellari M., Copin Y., 2003, MNRAS, 342, 345

\bibitem[{Cappellari \& Emsellem(2004)}]{Cappellari:2004p3283}
Cappellari M., Emsellem E., 2004, PASP, 116, 138

\bibitem[{{Cappellari} {et~al}\mbox{.}(2011{\natexlab{a}}){Cappellari},
  {Emsellem}, {Krajnovi{\'c}}, {McDermid}, {Scott}, {Verdoes Kleijn}, {Young},
  {Alatalo}, {Bacon}, \& {Blitz}}]{Cap2010}
{Cappellari} M. {et~al.}, 2011{\natexlab{a}}, MNRAS, 413, 813, ({P}aper I)

\bibitem[{{Cappellari} {et~al}\mbox{.}(2011{\natexlab{b}}){Cappellari},
  {Emsellem}, {Krajnovic}, {McDermid}, {Serra}, {Alatalo}, {Blitz}, {Bois},
  {Bournaud}, {Bureau}, \& {Davies}}]{Cap2011}
---, 2011{\natexlab{b}}, ArXiv e-prints 1104.3545, ({P}aper VII)

\bibitem[{{Clemens} {et~al}\mbox{.}(2010){Clemens}, {Jones}, {Bressan}, {Baes},
  {Bendo}, {Bianchi}, {Bomans}, {Boselli}, {Corbelli}, {Cortese}, {Dariush},
  {Davies}, {de Looze}, {di Serego Alighieri}, \& {Fadda}}]{Clemens:2010p3473}
{Clemens} M.~S. {et~al.}, 2010, A\&A, 518, L50

\bibitem[{Colbert, Mulchaey \& Zabludoff(2001)Colbert, Mulchaey, \&
  Zabludoff}]{Colbert:2001p1791}
Colbert J.~W., Mulchaey J.~S., Zabludoff A.~I., 2001, AJ, 121, 808

\bibitem[{{Combes}, {Young} \& {Bureau}(2007){Combes}, {Young}, \&
  {Bureau}}]{Combes:2007p231}
{Combes} F., {Young} L.~M., {Bureau} M., 2007, MNRAS, 377, 1795

\bibitem[{Crocker {et~al}\mbox{.}(2008)Crocker, Bureau, Young, \&
  Combes}]{Crocker:2008p946}
Crocker A.~F., Bureau M., Young L.~M., Combes F., 2008, MNRAS, 386, 1811

\bibitem[{{Crocker} {et~al}\mbox{.}(2011){Crocker}, {Bureau}, {Young}, \&
  {Combes}}]{Crocker:2011p3291}
{Crocker} A.~F., {Bureau} M., {Young} L.~M., {Combes} F., 2011, MNRAS, 410,
  1197

\bibitem[{Crocker {et~al}\mbox{.}(2009)Crocker, Jeong, Komugi, Combes, Bureau,
  Young, \& Yi}]{Crocker:2009p3262}
Crocker A.~F., Jeong H., Komugi S., Combes F., Bureau M., Young L.~M., Yi S.,
  2009, MNRAS, 393, 1255

\bibitem[{{Davis} {et~al}\mbox{.}(2011){Davis}, {Bureau}, {Young}, {Alatalo},
  {Blitz}, {Cappellari}, {Scott}, {Bois}, {Bournaud}, {Davies}, {de Zeeuw},
  {Emsellem}, {Khochfar}, {Krajnovic}, {Kuntschner}, {Lablanche}, {McDermid},
  {Morganti}, {Naab}, {Oosterloo}, {Sarzi}, {Serra}, \&
  {Weijmans}}]{Davis:2011p3472}
{Davis} T.~A. {et~al.}, 2011, MNRAS, 414, 2, 968, ({P}aper V)

\bibitem[{de~Zeeuw {et~al}\mbox{.}(2002)de~Zeeuw, Bureau, Emsellem, Bacon,
  Carollo, Copin, Davies, Kuntschner, Miller, Monnet, Peletier, \&
  Verolme}]{deZeeuw:2002p1496}
de~Zeeuw P.~T. {et~al.}, 2002, MNRAS, 329, 513

\bibitem[{{Deason} {et~al}\mbox{.}(2011){Deason}, {McCarthy}, {Font}, {Evans},
  {Frenk}, {Belokurov}, {Libeskind}, {Crain}, \& {Theuns}}]{Deason:2011p3441}
{Deason} A.~J. {et~al.}, 2011, ArXiv e-prints

\bibitem[{di~Serego~Alighieri {et~al}\mbox{.}(2007)di~Serego~Alighieri,
  Gavazzi, Giovanardi, Giovanelli, Grossi, Haynes, Kent, Koopmann, Pellegrini,
  Scodeggio, \& Trinchieri}]{diSeregoAlighieri:2007p3399}
di~Serego~Alighieri S. {et~al.}, 2007, A\&A, 474, 851

\bibitem[{Dressler(1980)}]{Dressler:1980p2456}
Dressler A., 1980, ApJ, 236, 351

\bibitem[{Duc {et~al}\mbox{.}(2007)Duc, Braine, Lisenfeld, Brinks, \&
  Boquien}]{Duc:2007p3378}
Duc P.-A., Braine J., Lisenfeld U., Brinks E., Boquien M., 2007, A\&A, 475, 187

\bibitem[{{Duley} \& {Williams}(1993)}]{Duley:1993p3469}
{Duley} W.~W., {Williams} D.~A., 1993, MNRAS, 260, 37

\bibitem[{{Eliche-Moral} {et~al}\mbox{.}(2010){Eliche-Moral},
  {Gonz{\'a}lez-Garc{\'{\i}}a}, {Balcells}, {Aguerri}, {Gallego}, \&
  {Zamorano}}]{ElicheMoral:2009p2918}
{Eliche-Moral} M.~C., {Gonz{\'a}lez-Garc{\'{\i}}a} A.~C., {Balcells} M.,
  {Aguerri} J.~A.~L., {Gallego} J., {Zamorano} J., 2010, in American Institute
  of Physics Conference Series, Vol. 1240, American Institute of Physics
  Conference Series, {V.~P.~Debattista \& C.~C.~Popescu}, ed., pp. 237--238

\bibitem[{{Emsellem} {et~al}\mbox{.}(2011){Emsellem}, {Cappellari},
  {Krajnovi{\'c}}, {Alatalo}, {Blitz}, {Bois}, {Bournaud}, {Bureau}, {Davies},
  {Davis}, {de Zeeuw}, {Khochfar}, {Kuntschner}, \& {Lablanche}}]{Emsellem2010}
{Emsellem} E. {et~al.}, 2011, MNRAS, 414, 2, 888, ({P}aper III)

\bibitem[{Emsellem {et~al}\mbox{.}(2004)Emsellem, Cappellari, Peletier,
  McDermid, Bacon, Bureau, Copin, Davies, Krajnovi{\'c}, Kuntschner, Miller, \&
  de~Zeeuw}]{Emsellem:2004p1497}
Emsellem E. {et~al.}, 2004, MNRAS, 352, 721

\bibitem[{{Faber} {et~al}\mbox{.}(2007){Faber}, {Willmer}, {Wolf}, {Koo},
  {Weiner}, {Newman}, {Im}, {Coil}, {Conroy}, {Cooper}, {Davis}, {Finkbeiner},
  {Gerke}, {Gebhardt}, {Groth}, {Guhathakurta}, {Harker}, {Kaiser}, \&
  {Kassin}}]{Faber:2007p3453}
{Faber} S.~M. {et~al.}, 2007, ApJ, 665, 265

\bibitem[{Forman, Jones \& Tucker(1985)Forman, Jones, \&
  Tucker}]{Forman:1985p3426}
Forman W., Jones C., Tucker W., 1985, ApJ, 293, 102

\bibitem[{{Garc{\'{\i}}a-Ruiz}, {Sancisi} \&
  {Kuijken}(2002){Garc{\'{\i}}a-Ruiz}, {Sancisi}, \&
  {Kuijken}}]{GarciaRuiz:2002p3468}
{Garc{\'{\i}}a-Ruiz} I., {Sancisi} R., {Kuijken} K., 2002, AAP, 394, 769

\bibitem[{{Giovanardi}, {Krumm} \& {Salpeter}(1983){Giovanardi}, {Krumm}, \&
  {Salpeter}}]{Giovanardi:1983p3439}
{Giovanardi} C., {Krumm} N., {Salpeter} E.~E., 1983, AJ, 88, 1719

\bibitem[{Giovanelli \& Haynes(1983)}]{Giovanelli:1983p3254}
Giovanelli R., Haynes M.~P., 1983, AJ, 88, 881

\bibitem[{Giovanelli {et~al}\mbox{.}(2005)Giovanelli, Haynes, Kent, Perillat,
  Saintonge, Brosch, Catinella, Hoffman, Stierwalt, Spekkens, Lerner, Masters,
  Momjian, Rosenberg, \& Springob}]{Giovanelli:2005p3400}
Giovanelli R. {et~al.}, 2005, AJ, 130, 2598

\bibitem[{{Grossi} {et~al}\mbox{.}(2009){Grossi}, {di Serego Alighieri},
  {Giovanardi}, {Gavazzi}, {Giovanelli}, {Haynes}, {Kent}, {Pellegrini},
  {Stierwalt}, \& {Trinchieri}}]{Grossi:2009p3448}
{Grossi} M. {et~al.}, 2009, A\&A, 498, 407

\bibitem[{Gunn \& Gott(1972)}]{Gunn:1972p3253}
Gunn J.~E., Gott J.~R., 1972, ApJ, 176, 1

\bibitem[{Heitsch \& Hartmann(2008)}]{Heitsch:2008p3258}
Heitsch F., Hartmann L., 2008, ApJ, 689, 290

\bibitem[{Helsdon \& Ponman(2003)}]{Helsdon:2003p3363}
Helsdon S.~F., Ponman T.~J., 2003, MNRAS, 339, L29

\bibitem[{Jungwiert, Combes \& Palou{\v s}(2001)Jungwiert, Combes, \& Palou{\v
  s}}]{Jungwiert:2001p3209}
Jungwiert B., Combes F., Palou{\v s} J., 2001, A\&A, 376, 85

\bibitem[{{Kannappan} \& {Fabricant}(2001)}]{Kannappan:2001p3451}
{Kannappan} S.~J., {Fabricant} D.~G., 2001, AJ, 121, 140

\bibitem[{{Kannappan}, {Guie} \& {Baker}(2009){Kannappan}, {Guie}, \&
  {Baker}}]{Kannappan:2009p3436}
{Kannappan} S.~J., {Guie} J.~M., {Baker} A.~J., 2009, AJ, 138, 579

\bibitem[{Kautsch {et~al}\mbox{.}(2008)Kautsch, Gonzalez, Soto, Tran, Zaritsky,
  \& Moustakas}]{Kautsch:2008p3360}
Kautsch S.~J., Gonzalez A.~H., Soto C.~A., Tran K.-V.~H., Zaritsky D.,
  Moustakas J., 2008, ApJ, 688, L5

\bibitem[{Kaviraj {et~al}\mbox{.}(2007)Kaviraj, Schawinski, Devriendt,
  Ferreras, Khochfar, Yoon, Yi, Deharveng, Boselli, Barlow, Conrow, Forster,
  Friedman, \& Martin}]{Kaviraj:2007p1804}
Kaviraj S. {et~al.}, 2007, ApJS, 173, 619

\bibitem[{Kenney \& Young(1986)}]{Kenney:1986p2835}
Kenney J.~D., Young J.~S., 1986, ApJ, 301, L13

\bibitem[{Kennicutt (1998)}]{Kennicutt:1998p3214}
Kennicutt R.~C., 1998, ApJ v.498, 498, 541

\bibitem[{Khochfar {et~al}\mbox{.}, submitted}]{Khochfar2011}
Khochfar S., submitted MNRAS, ({P}aper VIII)


\bibitem[{Knapp \& Rupen(1996)}]{Knapp:1996p1859}
Knapp G.~R., Rupen M.~P., 1996, ApJ v.460, 460, 271

\bibitem[{{Knapp}, {Turner} \& {Cunniffe}(1985){Knapp}, {Turner}, \&
  {Cunniffe}}]{Knapp:1985p3440}
{Knapp} G.~R., {Turner} E.~L., {Cunniffe} P.~E., 1985, AJ, 90, 454

\bibitem[{Kodama {et~al}\mbox{.}(2001)Kodama, Smail, Nakata, Okamura, \&
  Bower}]{Kodama:2001p3362}
Kodama T., Smail I., Nakata F., Okamura S., Bower R.~G., 2001, ApJ, 562, L9

\bibitem[{Krajnovi{\'c} {et~al}\mbox{.}(2008)Krajnovi{\'c}, Bacon, Cappellari,
  Davies, de~Zeeuw, Emsellem, Falc{\'o}n-Barroso, Kuntschner, McDermid,
  Peletier, Sarzi, van~den Bosch, \& van~de Ven}]{Krajnovic:2008p1491}
Krajnovi{\'c} D. {et~al.}, 2008, MNRAS, 390, 93

\bibitem[{Krajnovi{\'c} {et~al}\mbox{.}(2006)Krajnovi{\'c}, Cappellari,
  de~Zeeuw, \& Copin}]{Krajnovic:2006p2929}
Krajnovi{\'c} D., Cappellari M., de~Zeeuw P.~T., Copin Y., 2006, MNRAS, 366,
  787

\bibitem[{{Krajnovic} {et~al}\mbox{.}(2011){Krajnovic}, {Emsellem},
  {Cappellari}, {Alatalo}, {Blitz}, {Bois}, {Bournaud}, {Bureau}, {Davies},
  {Davis}, \& {de Zeeuw}}]{Kraj2010}
{Krajnovic} D. {et~al.}, 2011, MNRAS, online early, ({P}aper II)

\bibitem[{Krumm, van Driel \& van Woerden(1985)Krumm, van Driel, \& van
  Woerden}]{Krumm:1985p3248}
Krumm N., van Driel W., van Woerden H., 1985, A\&A, 144, 202

\bibitem[{{Kuntschner} {et~al}\mbox{.}(2006){Kuntschner}, {Emsellem}, {Bacon},
  {Bureau}, {Cappellari}, {Davies}, {de Zeeuw}, {Falc{\'o}n-Barroso},
  {Krajnovi{\'c}}, {McDermid}, {Peletier}, \& {Sarzi}}]{Kuntschner:2006p1489}
{Kuntschner} H. {et~al.}, 2006, MNRAS, 369, 497

\bibitem[{{Kuntschner} {et~al}\mbox{.}(2010){Kuntschner}, {Emsellem}, {Bacon},
  {Cappellari}, {Davies}, {de Zeeuw}, {Falc{\'o}n-Barroso}, {Krajnovi{\'c}},
  {McDermid}, {Peletier}, {Sarzi}, {Shapiro}, {van den Bosch}, \& {van de
  Ven}}]{Kuntschner:2010p3471}
---, 2010, MNRAS, 408, 97

\bibitem[{Li \& Seaquist(1994)}]{Li:1994p3359}
Li J.~G., Seaquist E.~R., 1994, AJ, 107, 1953

\bibitem[{Lia, Portinari \& Carraro(2002)Lia, Portinari, \&
  Carraro}]{Lia:2002p3210}
Lia C., Portinari L., Carraro G., 2002, MNRAS, 330, 821

\bibitem[{{Martig} \& {Bournaud}(2010)}]{Martig:2010p3475}
{Martig} M., {Bournaud} F., 2010, ApJL, 714, L275

\bibitem[{Martig {et~al}\mbox{.}(2009)Martig, Bournaud, Teyssier, \&
  Dekel}]{Martig:2009p2923}
Martig M., Bournaud F., Teyssier R., Dekel A., 2009, ApJ, 707, 250

\bibitem[{Mazzuca {et~al}\mbox{.}(2006)Mazzuca, Sarzi, Knapen, Veilleux, \&
  Swaters}]{Mazzuca:2006p3379}
Mazzuca L.~M., Sarzi M., Knapen J.~H., Veilleux S., Swaters R., 2006, ApJ, 649,
  L79

\bibitem[{McDermid {et~al}\mbox{.}(2006)McDermid, Emsellem, Shapiro, Bacon,
  Bureau, Cappellari, Davies, de~Zeeuw, Falc{\'o}n-Barroso, Krajnovi{\'c},
  Kuntschner, Peletier, \& Sarzi}]{McDermid:2006p1493}
McDermid R.~M. {et~al.}, 2006, MNRAS, 373, 906

\bibitem[{Morganti {et~al}\mbox{.}(2006)Morganti, de~Zeeuw, Oosterloo,
  McDermid, Krajnovi{\'c}, Cappellari, Kenn, Weijmans, \&
  Sarzi}]{Morganti:2006p1934}
Morganti R. {et~al.}, 2006, MNRAS, 371, 157

\bibitem[{Oosterloo {et~al}\mbox{.}(2010)Oosterloo, Morganti, Crocker,
  J{\"u}tte, Cappellari, de~Zeeuw, Krajnovi{\'c}, McDermid, Kuntschner, Sarzi,
  \& Weijmans}]{Oosterloo:2010p3376}
Oosterloo T. {et~al.}, 2010, MNRAS, 1397

\bibitem[{O'Sullivan, Forbes \& Ponman(2001)O'Sullivan, Forbes, \&
  Ponman}]{OSullivan:2001p3380}
O'Sullivan E., Forbes D.~A., Ponman T.~J., 2001, MNRAS, 328, 461

\bibitem[{Parriott \& Bregman(2008)}]{Parriott:2008p2869}
Parriott J.~R., Bregman J.~N., 2008, ApJ, 681, 1215

\bibitem[{Pozzetti {et~al}\mbox{.}(2007)Pozzetti, Bolzonella, Lamareille,
  Zamorani, Franzetti, F{\`e}vre, Iovino, Temporin, Ilbert, Arnouts, Charlot,
  Brinchmann, Zucca, Tresse, \& Scodeggio}]{Pozzetti:2007p3211}
Pozzetti L. {et~al.}, 2007, A\&A, 474, 443

\bibitem[{Regan {et~al}\mbox{.}(2001)Regan, Thornley, Helfer, Sheth, Wong,
  Vogel, Blitz, \& Bock}]{Regan:2001p3275}
Regan M.~W., Thornley M.~D., Helfer T.~T., Sheth K., Wong T., Vogel S.~N.,
  Blitz L., Bock D. C.-J., 2001, ApJ, 561, 218

\bibitem[{Romer {et~al}\mbox{.}(2000)Romer, Nichol, Holden, Ulmer, Pildis,
  Merrelli, Adami, Burke, Collins, Metevier, Kron, \&
  Commons}]{Romer:2000p3365}
Romer A.~K. {et~al.}, 2000, ApJS, 126, 209

\bibitem[{Roscoe(1999)}]{Roscoe:1999p3276}
Roscoe D.~F., 1999, A\&A, 343, 788

\bibitem[{{Sage} \& {Welch}(2006)}]{Sage:2006p3466}
{Sage} L.~J., {Welch} G.~A., 2006, ApJ, 644, 850

\bibitem[{{Sage}, {Welch} \& {Young}(2007){Sage}, {Welch}, \&
  {Young}}]{Sage:2007p3467}
{Sage} L.~J., {Welch} G.~A., {Young} L.~M., 2007, ApJ, 657, 232

\bibitem[{{Sarzi} {et~al}\mbox{.}(2007){Sarzi}, {Bacon}, {Cappellari},
  {Davies}, {Emsellem}, {Falc{\'o}n-Barroso}, {Krajnovi{\'c}}, {Kuntschner},
  {McDermid}, {Peletier}, {de Zeeuw}, \& {van de Ven}}]{Sarzi:2007p3289}
{Sarzi} M. {et~al.}, 2007, New Astronomy Reviews, 51, 18

\bibitem[{Sarzi {et~al}\mbox{.}(2006)Sarzi, Falc{\'o}n-Barroso, Davies, Bacon,
  Bureau, Cappellari, de~Zeeuw, Emsellem, Fathi, Krajnovi{\'c}, Kuntschner,
  McDermid, \& Peletier}]{Sarzi:2006p1474}
Sarzi M. {et~al.}, 2006, MNRAS, 366, 1151

\bibitem[{{Sarzi} {et~al}\mbox{.}(2010){Sarzi}, {Shields}, {Schawinski},
  {Jeong}, {Shapiro}, {Bacon}, {Bureau}, {Cappellari}, {Davies}, {de Zeeuw},
  {Emsellem}, {Falc{\'o}n-Barroso}, {Krajnovi{\'c}}, {Kuntschner}, {McDermid},
  {Peletier}, {van den Bosch}, {van de Ven}, \& {Yi}}]{Sarzi:2009p2933}
{Sarzi} M. {et~al.}, 2010, MNRAS, 402, 2187

\bibitem[{Sault, Teuben \& Wright(1995)Sault, Teuben, \&
  Wright}]{Sault:1995p2768}
Sault R.~J., Teuben P.~J., Wright M. C.~H., 1995, Astronomical Data Analysis
  Software and Systems IV, 77, 433

\bibitem[{Schinnerer \& Scoville(2002)}]{Schinnerer:2002p981}
Schinnerer E., Scoville N., 2002, ApJ, 577, L103

\bibitem[{{Serra} {et~al}\mbox{.}(2009){Serra}, {Morganti}, {Oosterloo},
  {Alatalo}, {Blitz}, {Bois}, {van den Bosch}, {Bournaud}, {Bureau},
  {Cappellary}, {Davies}, {Davis}, {Duc}, {Emsellem}, \&
  {Falcon-Barroso}}]{Serraconfprop}
{Serra} P. {et~al.}, 2009, in Panoramic Radio Astronomy: Wide-field 1-2 GHz
  Research on Galaxy Evolution

\bibitem[{Shapiro {et~al}\mbox{.}(2010)Shapiro, Falc{\'o}n-Barroso, van~de Ven,
  de~Zeeuw, Sarzi, Bacon, Bolatto, Cappellari, Croton, Davies, Emsellem,
  Fakhouri, Krajnovi{\'c}, Kuntschner, McDermid, Peletier, van~den Bosch, \&
  van~der Wolk}]{Shapiro:2010p2932}
Shapiro K.~L. {et~al.}, 2010, MNRAS, 402, 2140

\bibitem[{{Skillman} {et~al}\mbox{.}(1996){Skillman}, {Kennicutt}, {Shields},
  \& {Zaritsky}}]{Skillman:1996p3470}
{Skillman} E.~D., {Kennicutt}, Jr. R.~C., {Shields} G.~A., {Zaritsky} D., 1996,
  ApJ, 462, 147

\bibitem[{Springel, Matteo \& Hernquist(2005)Springel, Matteo, \&
  Hernquist}]{Springel:2005p3425}
Springel V., Matteo T.~D., Hernquist L., 2005, ApJ, 620, L79

\bibitem[{Toomre(1964)}]{Toomre:1964p3272}
Toomre A., 1964, ApJ, 139, 1217

\bibitem[{van Dokkum {et~al}\mbox{.}(1999)van Dokkum, Franx, Fabricant, Kelson,
  \& Illingworth}]{vanDokkum:1999p3277}
van Dokkum P.~G., Franx M., Fabricant D., Kelson D.~D., Illingworth G.~D.,
  1999, ApJ, 520, L95

\bibitem[{{Verheijen} \& {Zwaan}(2001)}]{Verheijen:2001p3449}
{Verheijen} M.~A.~W., {Zwaan} M., 2001, in Astronomical Society of the Pacific
  Conference Series, Vol. 240, Gas and Galaxy Evolution, {J.~E.~Hibbard,
  M.~Rupen, \& J.~H.~van Gorkom}, ed., pp. 867--+

\bibitem[{Vollmer, Huchtmeier \& van Driel(2005)Vollmer, Huchtmeier, \& van
  Driel}]{Vollmer:2005p3247}
Vollmer B., Huchtmeier W., van Driel W., 2005, A\&A, 439, 921

\bibitem[{{Wada} \& {Habe}(1995)}]{Wada:1995p3443}
{Wada} K., {Habe} A., 1995, MNRAS, 277, 433

\bibitem[{{Welch} \& {Sage}(2003)}]{Welch:2003p2521}
{Welch} G.~A., {Sage} L.~J., 2003, ApJ, 584, 260

\bibitem[{{Welch}, {Sage} \& {Young}(2010){Welch}, {Sage}, \&
  {Young}}]{Welch:2010p3290}
{Welch} G.~A., {Sage} L.~J., {Young} L.~M., 2010, ApJ, 725, 100

\bibitem[{{Yi}(2008)}]{Yi:2008p3435}
{Yi} S.~K., 2008, in Astronomical Society of the Pacific Conference Series,
  Vol. 392, Hot Subdwarf Stars and Related Objects, {U.~Heber, C.~S.~Jeffery,
  \& R.~Napiwotzki}, ed., pp. 3--+

\bibitem[{{Yi} {et~al}\mbox{.}(2005){Yi}, {Yoon}, {Kaviraj}, {Deharveng},
  {Rich}, {Salim}, {Boselli}, {Lee}, {Ree}, {Sohn}, {Rey}, {Lee}, {Rhee},
  {Bianchi}, {Byun}, {Donas}, {Friedman}, {Heckman}, {Jelinsky}, {Madore},
  {Malina}, \& {Martin}}]{Yi:2005p3450}
{Yi} S.~K. {et~al.}, 2005, ApJL, 619, L111

\bibitem[{Young(2002)}]{Young:2002p943}
Young L.~M., 2002, AJ, 124, 788

\bibitem[{Young, Bureau \& Cappellari(2008)Young, Bureau, \&
  Cappellari}]{Young:2008p788}
Young L.~M., Bureau M., Cappellari M., 2008, ApJ, 676, 317

\bibitem[{{Young} {et~al}\mbox{.}(2011){Young}, {Bureau}, {Davis}, {Combes},
  {McDermid}, {Alatalo}, {Blitz}, {Bois}, {Bournaud}, {Cappellari}, \&
  {Davies}}]{Young2010}
{Young} L.~M. {et~al.}, 2011, MNRAS, 414, 2, 940 ({P}aper IV)

\bibitem[{Zabludoff \& Mulchaey(1998)}]{Zabludoff:1998p3361}
Zabludoff A.~I., Mulchaey J.~S., 1998, ApJ v.496, 496, 39

\end{thebibliography}

\label{lastpage}

\begin{table*}
\caption{Kinematic misalignment between the ionised gas and the stars for all the detected, fast-rotating, \atlas\ early-type galaxies.}
\begin{tabular*}{0.5\textwidth}{@{\extracolsep{\fill}}r r r r r r r r c}
\hline
Name & $\phi_{\rm ion}$ & $\Delta\phi_{\rm ion}$ & $\Psi_{\rm ion-star}$ & $\Delta\Psi_{\rm ion-star}$ \\
& (deg) & (deg) & (deg) & (deg) \\
(1) & (2) & (3) & (4) & (5)\\
\hline
IC0676 & 348.5 & 26.9 & 30.5 & 29.6\\
IC0719 & 232.5 & 2.5 & 173.5 & 26.9\\
IC1024 & 31.5 & 8.2 & 2.0 & 13.6\\
NGC0680 & 26.0 & 2.5 & 26.5 & \hspace{4pt}4.5\\
NGC1266 & 350.5 & 3.5 & 56.0 & \hspace{4pt}7.8\\
NGC2577 & 88.0 & 2.8 & 16.0 & \hspace{4pt}3.4\\
NGC2764 & 189.5 & 5.2 & 6.5 & \hspace{4pt}8.6\\
NGC2778 & 50.5 & 4.0 & 5.0 & \hspace{4pt}6.2\\
NGC2824 & 159.5 & 7.5 & 0.0 & \hspace{4pt}8.0\\
NGC2852 & 188.0 & 2.8 & 32.0 & \hspace{4pt}5.5\\
NGC2859 & 269.0 & 1.8 & 5.0 & \hspace{4pt}3.5\\
NGC2950 & 97.5 & 2.2 & 16.5 & \hspace{4pt}4.0\\
NGC3182 & 316.5 & 6.0 & 4.0 & \hspace{4pt}9.2\\
NGC3226 & 36.0 & 1.0 & 8.0 & \hspace{4pt}6.3\\
NGC3245 & 184.0 & 4.8 & 9.5 & \hspace{4pt}5.6\\
NGC3248 & 302.5 & 4.0 & 179.5 & \hspace{4pt}9.8\\
NGC3412 & 175.0 & 10.5 & 19.0 & 12.1\\
NGC3457 & 339.5 & 4.8 & 5.5 & 38.5\\
NGC3499 & 125.0 & 4.2 & 75.0 & 12.7\\
NGC3595 & 99.5 & 7.2 & 99.0 & \hspace{4pt}7.4\\
NGC3599 & 352.5 & 3.2 & 63.0 & 17.6\\
NGC3607 & 302.5 & 0.5 & 1.0 & \hspace{4pt}2.8\\
NGC3610 & 128.0 & 3.8 & 6.5 & \hspace{4pt}3.8\\
NGC3613 & 96.0 & 0.5 & 3.5 & \hspace{4pt}1.3\\
NGC3619 & 76.0 & 0.5 & 24.0 & \hspace{4pt}3.0\\
NGC3626 & 165.0 & 3.2 & 174.5 & \hspace{4pt}4.6\\
NGC3648 & 315.0 & 5.2 & 60.5 & \hspace{4pt}6.0\\
NGC3665 & 209.5 & 2.2 & 4.0 & \hspace{4pt}3.0\\
NGC3694 & 309.5 & 8.0 & 159.5 & 12.2\\
NGC3838 & 137.0 & 5.2 & 1.5 & \hspace{4pt}6.3\\
NGC3941 & 144.0 & 1.2 & 129.0 & \hspace{4pt}3.7\\
NGC4026 & 343.0 & 2.0 & 18.5 & \hspace{4pt}3.6\\
NGC4036 & 258.0 & 1.5 & 3.0 & \hspace{4pt}1.8\\
NGC4111 & 197.0 & 1.0 & 47.5 & \hspace{4pt}2.5\\
NGC4119 & 292.5 & 18.6 & 1.0 & 19.7\\
NGC4143 & 155.0 & 1.8 & 165.5 & \hspace{4pt}2.9\\
NGC4179 & 142.8 & 3.1 & 0.7 & \hspace{4pt}4.4\\
NGC4203 & 198.0 & 1.8 & 3.5 & \hspace{4pt}6.0\\
NGC4251 & 207.0 & 1.2 & 71.5 & \hspace{4pt}3.2\\
NGC4255 & 113.0 & 2.8 & 2.0 & \hspace{4pt}4.3\\
NGC4281 & 118.0 & 0.5 & 33.0 & \hspace{4pt}1.6\\
NGC4324 & 239.0 & 6.8 & 1.0 & \hspace{4pt}8.6\\
NGC4382 & 29.0 & 9.7 & 8.5 & 10.8\\
NGC4417 & 224.0 & 9.5 & 4.5 & 10.4\\
NGC4429 & 92.5 & 2.0 & 6.0 & \hspace{4pt}3.2\\
NGC4435 & 198.6 & 3.1 & 6.1 & \hspace{4pt}3.6\\
NGC4473 & 101.0 & 9.7 & 9.0 & 10.5\\
NGC4474 & 95.5 & 4.5 & 16.5 & \hspace{4pt}8.1\\
NGC4494 & 186.2 & 3.1 & 1.2 & \hspace{4pt}6.8\\
NGC4521 & 12.5 & 1.8 & 23.5 & \hspace{4pt}3.3\\
NGC4596 & 112.5 & 0.5 & 12.5 & \hspace{4pt}4.5\\
NGC4612 & 310.3 & 3.1 & 17.7 & 12.0\\
NGC4621 & 307.0 & 10.0 & 14.0 & 11.2\\
NGC4643 & 68.0 & 2.5 & 20.0 & \hspace{4pt}4.9\\
NGC4684 & 292.0 & 7.5 & 87.5 & \hspace{4pt}9.0\\
NGC4694 & 167.0 & 45.0 & 157.5 & 48.9\\
NGC4697 & 218.0 & 0.5 & 29.5 & \hspace{4pt}2.1\\
NGC4710 & 208.5 & 2.2 & 1.0 & \hspace{4pt}4.4\\
NGC4753 & 85.5 & 1.5 & 3.0 & \hspace{4pt}2.9\\
NGC5103 & 357.0 & 8.0 & 38.5 & \hspace{4pt}8.9\\
NGC5173 & 93.5 & 6.0 & 174.0 & 17.8\\
NGC5273 & 71.5 & 44.5 & 119.0 & 45.0\\
NGC5353 & 297.5 & 0.5 & 24.5 & \hspace{4pt}1.1\\
NGC5355 & 62.0 & 8.5 & 33.0 & 16.4\\
NGC5379 & 77.0 & 5.0 & 16.0 & 11.2\\
\hline
\end{tabular*}
\parbox[t]{0.5 \textwidth}{Table continued below.}
\label{datatable1}
\end{table*}

\begin{table*}
\begin{tabular*}{0.5\textwidth}{@{\extracolsep{\fill}}r r r r r r r r c}
\hline
Name & $\phi_{\rm ion}$ & $\Delta\phi_{\rm ion}$ & $\Psi_{\rm ion-star}$ & $\Delta\Psi_{\rm ion-star}$ \\
& (deg) & (deg) & (deg) & (deg) \\

(1) & (2) & (3) & (4) & (5)\\
\hline
NGC5422 & 348.5 & 2.0 & 14.5 & \hspace{4pt}4.2\\
NGC5485 & 264.0 & 4.2 & 5.0 & \hspace{4pt}8.0\\
NGC5493 & 137.5 & 5.2 & 16.0 & \hspace{4pt}6.5\\
NGC5582 & 23.5 & 2.5 & 6.0 & \hspace{4pt}3.7\\
NGC5866 & 126.5 & 3.2 & 0.0 & \hspace{4pt}3.5\\
NGC6014 & 151.5 & 10.2 & 4.5 & 13.3\\
NGC6017 & 132.5 & 5.2 & 0.0 & \hspace{4pt}8.4\\
NGC6149 & 218.0 & 7.5 & 17.0 & \hspace{4pt}8.6\\
NGC6278 & 337.0 & 3.0 & 31.5 & \hspace{4pt}5.6\\
NGC6798 & 313.0 & 1.2 & 174.0 & \hspace{4pt}6.9\\
NGC7465 & 109.5 & 9.5 & 57.0 & 30.5\\
NGC7710 & 154.0 & 21.2 & 20.0 & 46.6\\
PGC016060 & 161.5 & 2.8 & 2.5 & \hspace{4pt}3.4\\
PGC029321 & 105.5 & 27.2 & 55.0 & 46.0\\
PGC035754 & 47.0 & 12.0 & 39.0 & 16.3\\
PGC042549 & 242.0 & 5.0 & 1.0 & \hspace{4pt}7.8\\
PGC056772 & 214.0 & 2.8 & 23.0 & \hspace{4pt}6.1\\
PGC058114 & 93.1 & 3.1 & 153.9 & 10.5\\
PGC061468 & 73.0 & 15.2 & 32.0 & 19.1\\
UGC05408 & 122.5 & 10.2 & 27.5 & 19.9\\
UGC06176 & 214.0 & 8.5 & 13.5 & 10.7\\
UGC09519 & 172.0 & 4.5 & 77.5 & \hspace{4pt}6.2\\
\hline
\end{tabular*}
\parbox[t]{0.5 \textwidth}{ \textit{Notes:} Columns 2-5 show the ionised gas kinematic PA, error, and the misalignments between the ionised gas and the main body of the stars for  all the fast-rotating \atlas\ galaxies where a kinematic PA was measurable. The stellar PA is taken directly from Paper II, and the data tables from this paper are available to download from http://www.purl.org/atlas3d. Kinematic misalignments for the fast-rotating SAURON ETGs are tabulated in \cite{Sarzi:2006p1474}.}
\end{table*}

\begin{table*}
\caption{Kinematic misalignment between the ionised gas and the stars for all the slow-rotating \atlas\ early-type galaxies with measurable kinematic misalignments.}
\begin{tabular*}{0.5\textwidth}{@{\extracolsep{\fill}}r r r r r r r r c}
\hline
Name & $\phi_{\rm ion}$ & $\Delta\phi_{\rm ion}$ & $\Psi_{\rm ion-star}$ & $\Delta\Psi_{\rm ion-star}$ \\
& (deg) & (deg) & (deg) & (deg) \\
(1) & (2) & (3) & (4) & (5)\\
\hline
NGC0661 & 227.5 & 2.8 & 8.5 & 12.3\\
NGC1222 & 54.5 & 21.5 & 11.5 & 23.5\\
NGC1289 & 276.0 & 15.0 & 176.0 & 18.0\\
NGC3522 & 192.0 & 5.5 & 78.5 & 89.9\\
NGC3796 & 67.5 & 12.2 & 58.0 & 18.8\\
NGC4168 & 102.5 & 1.2 & 142.5 & 89.8\\
NGC4191 & 178.0 & 5.7 & 4.5 & 7.5\\
NGC4261 & 103.0 & 0.5 & 52.0 & 3.3\\
NGC4476 & 219.0 & 7.5 & 12.5 & 13.7\\
NGC4636 & 32.5 & 1.0 & 125.5 & 89.8\\
NGC4690 & 191.5 & 1.5 & 139.5 & 25.8\\
NGC5322 & 102.0 & 1.0 & 171.0 & 7.3\\
NGC5481 & 233.5 & 3.8 & 7.5 & 19.4\\
NGC5631 & 305.0 & 2.0 & 174.0 & 9.0\\
PGC028887 & 216.5 & 8.0 & 4.5 & 12.4\\
\hline
\end{tabular*}
\parbox[t]{0.5 \textwidth}{ \textit{Notes:} Columns defined as in Table \protect \ref{datatable1}. Kinematic misalignments for the slow-rotating SAURON ETGs are tabulated in \cite{Sarzi:2006p1474}.}
\label{srtable}
\end{table*}

\end{document}